\documentclass[fleqn,usenatbib]{mnras}

\usepackage[T1]{fontenc}

\DeclareRobustCommand{\VAN}[3]{#2}
\let\VANthebibliography\thebibliography
\def\thebibliography{\DeclareRobustCommand{\VAN}[3]{##3}\VANthebibliography}

\usepackage{array,multirow,graphicx} % Including figure files
\usepackage{float} 
\usepackage{amsmath}  % Advanced maths commands
\usepackage{amssymb}  % Extra maths symbols
\usepackage{physics}
\usepackage{tabularx}
\usepackage{xcolor}
\usepackage{tikz}
\usepackage{tablefootnote}
\usepackage{comment}
\definecolor{lime}{HTML}{A6CE39}
\DeclareRobustCommand{\orcidicon}{%
  \begin{tikzpicture}
  \draw[lime, fill=lime] (0,0) 
  circle [radius=0.16] 
  node[white] {{\fontfamily{qag}\selectfont \tiny ID}};
  \draw[white, fill=white] (-0.0625,0.095) 
  circle [radius=0.007];
  \end{tikzpicture}
  \hspace{-2mm}
}

\foreach \x in {A, ..., Z}{%
  \expandafter\xdef\csname orcid\x\endcsname{\noexpand\href{https://orcid.org/\csname orcidauthor\x\endcsname}{\noexpand\orcidicon}}
}

\usepackage{newtxtext,newtxmath}
\newcommand\REPLY[1]{\textcolor{black}{#1}}
\title[VDF from gal-gal lenses]{Velocity dispersion function evolution from strong lensing statistics}

\author[G. Ferrami \& J.S.B. Wyithe]{
G. Ferrami $^{1,}$$^{2}$ \thanks{E-mail: gferrami@student.unimelb.edu.au } \orcidA{}
J. Stuart B. Wyithe $^{2,}$$^{3}$ \orcidB{}\\
$^{1}$School of Physics, University of Melbourne, Parkville, VIC 3010, Australia\\
$^{2}$ARC Centre of Excellence for All-Sky Astrophysics in 3 Dimensions (ASTRO 3D)\\
$^{3}$Research School of Astronomy and Astrophysics, Australian National University, Canberra, ACT 2611, Australia}

\date{Accepted XXX. Received YYY; in original form ZZZ}

\pubyear{2024}

\begin{document}
\label{firstpage}
\pagerange{\pageref{firstpage}--\pageref{lastpage}}
\maketitle

% Abstract of the paper - 250 words (200 words for Letters)
\begin{abstract}
The redshift and size distributions of galaxy scale strong lenses depend on the evolution of early-type galaxies (ETGs).
We use this dependence to constrain the velocity dispersion function (VDF) evolution from the Strong Lensing Legacy Survey (SL2S) sample of lenses \REPLY{in the redshift range $0.25\lesssim z \lesssim 0.75$}.
Our modeling of the lens population includes lens identifiability given survey parameters, and constrains the evolution of the VDF based on the redshift distributions of sources and lenses as well as the distribution of Einstein radii.
We consider five different assumptions for the reference VDF at redshift zero and two sets of scaling relations for the VDF. 
We find that in all cases the observed lens sample favors a slow evolution of both the VDF normalization factor and the VDF characteristic velocity with redshift which is consistent with a VDF that is constant in redshift for \REPLY{$z\lesssim0.75$}.
% The sign of the redshift evolution parameters depends on the fiducial profile assumed for the local stellar VDF. 
% We account for the spread in the relation between the velocity dispersion entering an isothermal lensing model and the central stellar velocity dispersion.
% Our best fit results are compatible with directly measured VDF profiles at redshift $0.5\lesssim z \lesssim1$, reconciling the tension between inferred and observed VDF profiles.
\end{abstract}


%%%%%%%%%%%%%%%%%%%%%%%%%%%%%%%%%%%%%%%%%%%%%%%%%%

%%%%%%%%%%%%%%%%% BODY OF PAPER %%%%%%%%%%%%%%%%%%

\section{Introduction}
The central stellar velocity dispersion measured in galaxies traces the depth of the gravitational potential well, the structure of the stellar orbits and the effects of stellar and AGN feedback. The distribution of velocity dispersion across cosmic time (i.e., the Velocity Dispersion Function, or VDF) is a key test for galaxy formation models.
The stellar velocity dispersion of local ETGs has been measured by the Sloan Digital Sky Survey (SDSS), with estimates of the associated VDF varying by a factor of up to a factor of $\sim3$ depending on the assumed completeness corrections (\citealt{Sheth_2003_VDF_from_LF}, \citealt{Mitchell_2005_VDF_from_LF}, \citealt{Choi_2007_SDSS}, \citealt{Chae_2010}, \citealt{Bernardi_VDF_2010}).
In the redshift interval $0.5 \lesssim z \lesssim 1$, the VDF profile has been directly measured by \cite{Montero_Dorta_VDF} and \cite{Taylor_VDF}, and has been inferred from photometry (\citealt{Bezanson_VDFevo}) or constrained from the local VDF using lensing (\citealt{Chae_2010}, \citealt{OguriVDF2012}, \citealt{Geng_2021}), the age distribution of local galaxies (\citealt{2009Shankar}) or the stellar mass function (\citealt{Mason_2015}).

However, the VDF evolution up to $z\sim1$ is largely unconstrained, as different studies tend do not qualitatively agree in the sign of the evolution, though most are consistent with a static distribution (e.g., see \citealt{Geng_2021}).
This is mainly due to the differences in the assumed local distribution of stellar velocity dispersion and the completeness function of the considered sample of lenses.

The statistical distribution of strong gravitational lenses depends on the mass function evolution of the deflector population and the cosmological distances between the observer, lens and source, and therefore probes both astrophysics and cosmology.
The total mass distribution of early-type galaxies (ETGs) is well approximated by an isothermal distribution within one Einstein radius, which is the typical radial scale of galaxy-scale strong lensing of order $\sim 1$ arcsec (e.g, \citealt{Gavazzi_SLACS_2007}, \citealt{Koopmans_bulge_halo_conspiracy}, \citealt{Lapi_2012}, \citealt{Sonnenfeld_SL2S_2013}, \citealt{SLACS_debiased}). 
Since a singular isothermal density profile depends only on the value of the associated velocity dispersion $\sigma$, it is natural to model the distribution of deflectors with the comoving density of galaxies per unit velocity dispersion (VDF).
Furthermore, in a sample of galaxy-scale lenses the value of the central stellar velocity dispersion $\sigma_\star$, directly measured from spectroscopy, tends to correlate linearly with the velocity dispersion $\sigma$ entering the isothermal mass profile that best fit the total mass distribution within its Einstein radius (e.g., \citealt{Kochanek_2000_fe}, \citealt{Treu_2006}, \citealt{Bolton_2008_fe}, \citealt{Grillo_2008_fe}, \citealt{Auger_2010_fe}, \citealt{Zahid_2018_fe}).
This allows an explicit relation to be established between the observed stellar VDF and the isothermal lens VDF associated to a population of lenses.\\
\indent Since most lens samples have redshifts of the deflectors between $0.2 \lesssim z_l \lesssim 1$, the distribution of lenses is affected by the evolution of the VDF in that redshift range.
Some of the techniques used to map lensing observables to the VDF evolution include the probability of observing a lensed background over random lines-of-sight, or optical depth to lensing (ODTL, \citealt{Turner_1984}), as employed in \citealt{Mitchell_2005_VDF_from_LF}; and the lens-redshift test, based on the differential lensing optical depth with respect to the angular Einstein radius (\citealt{Kochanek_1992}, \citealt{Ofek_2003}, \citealt{Capelo_Natarajan_2007}, \citealt{Geng_2021}).
To constrain the evolution of the VDF profile with a sample of lenses, it is common practice to anchor the local isothermal lens VDF at $z=0$ to a fiducial direct measurement of the central stellar velocity dispersion of ETGs, for a fixed cosmology.
% as well as fixing a background cosmology, as the evolution of the VDF and the Hubble parameter $H(z)$ are degenerate with respect to lensing observables.
Furthermore, an accurate estimation of the completeness of the Einstein radii distribution is critical to reliably constrain the evolution of the VDF profile, as demonstrated in \cite{Oguri2005} and \cite{Capelo_Natarajan_2007}.

In this paper we use a new strong lensing statistic model to study the evolution of the Velocity Dispersion Function of early-type galaxies, as a function of different fiducial local measurements of the VDF profile and accounting for the spread in the relation between the central stellar velocity dispersion and the velocity dispersion entering the isothermal total density profile. We study the lensing constraints on the VDF evolution using the SL2S sample of lenses (\citealt{Sonnenfeld_SL2S_2013a}).
We improve on previous analyses by:
\begin{itemize}
  \item including a lens model that can account for the details of identifiability of the lens observational sample (\citealt{Ferrami_Wyithe_Lens_stat_model}),
  \item correcting for the relation between $\sigma_\star$ and $\sigma$ so that the comparison with z$\sim$0 is self-consistent, 
  \item checking how sensitive the conclusions are to the systematics by comparing the results obtained from five fiducial local VDFs.
\end{itemize}

The outline of the paper is as follows. 
In Section 2, we briefly present the Velocity Dispersion Function and the different methods that can be used to constrain its evolution.
In Section 3, we introduce the sample of lenses used in this study.
In Section 4, we define the model that links the VDF distribution to the lensing observables.
In Section 5, we explain the method used to constrain the VDF evolution parameters, examine the results and compare them with previous studies.
Our conclusions are summarized in Section 6.
Throughout this paper, we fix the cosmology to be $H_0 = 70$ km s$^{-1}$ Mpc$^{-1}$, $\Omega_0 = 0.3$, $\Omega_\Lambda= 0.7$.

%######%######%######%######%######%######%######%######%######%######%######%######%######%

\section{Velocity Dispersion Function}

The Velocity Dispersion Function (VDF) is the comoving density per unit velocity of the central stellar velocity dispersion $\sigma_\star$ of a population of galaxies at a given redshift.
The VDF is therefore a tracer of the evolution of the underlying halo mass function and the dynamics and feedback in the inner regions of galaxies, as recorded by the relations between velocity dispersion and luminosity $L$-$\sigma_\star$ (\citealt{Faber_Jackson}, \citealt{Tully_Fisher}) or central black hole mass $M_{\rm{BH}}$-$\sigma_\star$ centr.

Since direct spectroscopic measurements of the stellar velocity dispersion require integration times that are too long to allow complete surveys of the whole galactic population beyond the local universe, a variety of techniques have been employed to map the redshift evolution of the VDF to other observables, such as the age of the stellar population (\citealt{2009Shankar}), photometric data (\citealt{Bezanson_VDFevo}), the stellar mass function (e.g., \citealt{Mason_2015}), or the statistical redshift and mass distribution of a sample of strong gravitational lenses (e.g. \citealt{Kochanek_1992}, \citealt{Mitchell_2005_VDF_from_LF}, \citealt{Capelo_Natarajan_2007}, \citealt{Chae_2010}, \citealt{Oguri_2012}, \citealt{Geng_2021}).
In the redshift interval $0.5 \lesssim z \lesssim 1$, the high-$\sigma_\star$ regime of the VDF profile has been directly measured by \cite{Montero_Dorta_VDF} and \cite{Taylor_VDF}.

In this paper, we aim to measure the VDF at $z\sim1$ and its evolution by constraining its change relative to $z=0$ using the distributions of observed lens redshifts and Einstein radii.
The functional form of the local VDF has been derived either from the luminosity function using a $L$-$\sigma_\star$ relation (Inferred Velocity Dispersion Function, \citealt{Kochanek_1992}), or directly fitting a modified Schechter profile to the Measured Velocity Dispersion Function (MVDF, \citealt{Sheth_2003_VDF_from_LF}).
These two approaches lead to incompatible estimates of the VDF in the high-$\sigma_\star$ and low-$\sigma_\star$ regime, mainly due to the scatter in the luminosity-$\sigma_\star$ relations (e.g., see \citealt{Capelo_Natarajan_2007} for a thorough discussion on the implications on strong lensing effects).
Throughout this work we adopt the MVDF in the form used in \cite{Capelo_Natarajan_2007},
\begin{equation}\label{eq:VDF}
\dv{\Phi}{\sigma_\star} = 
\frac{\beta}{\Gamma\left({\alpha}/{\beta}\right)}
\frac{\Phi_{\rm{s}}}{\sigma_{\rm{s}}}
\left(\frac{\sigma_\star}{\sigma_{\rm{s}}}\right)^{\alpha-1}
\exp\left[-\left(\frac{\sigma_\star}{\sigma_{\rm{s}}}\right)^{\beta}\right].
\end{equation}
At redshift $z\ll1$ the accretion of dark matter haloes follows a power law, while at high redshifts it is well described by an exponential (\citealt{Correa_2015}). The transition between the power-law and the exponential growth regimes happens at $0.6 \lesssim  z \lesssim  2$.
While the population of strong lensing deflectors is entirely at $z \lesssim 2$, most lie below $z < 0.6$ where the power law approximation of the halo growth holds. 
Since we expect the VDF to trace the halo virial velocity distribution evolution, we consider two sets of scaling relations for the redshift dependence of the halo VDF scale density $\Phi_{\rm{s}}$ and scale velocity $\sigma_{\rm{s}}$.
The first is a power law,
\begin{equation}\label{eq:scaling_power_law}
\begin{split}
\Phi_{\rm{s}}(z)   &= \Phi_{{\rm{s}},0} (1+z)^{\nu_n}\\
\sigma_{\rm{s}}(z) &= \sigma_{{\rm{s}},0} (1+z)^{\nu_v} \:,
\end{split}
\end{equation}
and the second is an exponential,
\begin{equation}\label{eq:scaling_exp}
\begin{split}
\Phi_{\rm{s}}(z)   &= \Phi_{{\rm{s}},0} 10^{Pz}\\
\sigma_{\rm{s}}(z) &= \sigma_{{\rm{s}},0} 10^{Uz} \:,
\end{split}
\end{equation}
where $\Phi_{{\rm{s}},0}$ and $\sigma_{{\rm{s}},0}$ are the characteristic values at redshift zero.

To test the sensitivity of the retrieved VDF evolution parameters on the fiducial galaxy velocity dispersion distribution in the local universe, we perform our analysis on five different reference VDFs at redshift $z=0$
(\citealt{Sheth_2003_VDF_from_LF}, \citealt{Mitchell_2005_VDF_from_LF}, \citealt{Choi_2007_SDSS}, \citealt{Chae_2010}, \citealt{Bernardi_VDF_2010}), with the values of the VDF parameters listed in Table \ref{tab:Fiducial_param}.
We include only estimates of the local VDF that are parametrized with a VDF profile as in Eq. (\ref{eq:VDF}) (non-parametric VDFs, like \citealt{Sohn_VDF_2017} based on field ETGs in SDSS DR12, are consistent with the VDF from \cite{Choi_2007_SDSS} for $\sigma \gtrsim 200$ km/s, which is roughly the scale at which lensing starts to become effective in the SL2S survey).

\begin{table}
  \caption{Fiducial parameters for the $z=0$ stellar VDF.}
  \label{tab:Fiducial_param}
  \centering
  \begin{tabular*}{\linewidth}{@{\extracolsep{\fill}} c c c c c}
      \hline 
      Reference & $\Phi_\star$ [Mpc$^{-3}$]& $\sigma_\star$ [km s$^{-1}$] & $\alpha$ & $\beta$\\
      \hline 
      \hline 
      \cite{Sheth_2003_VDF_from_LF}    & 2.0$\times10^{-3}$ & 88    & 6.5   & 1.93   \\
      \cite{Mitchell_2005_VDF_from_LF} & 1.4$\times10^{-3}$ & 88.8  & 6.5   & 1.93   \\
      \cite{Choi_2007_SDSS}            & 2.7$\times10^{-3}$ & 161   & 2.32  & 2.67   \\
      \cite{Chae_2010}                 & 4.46$\times10^{-3}$& 217   & 0.85  & 3.72   \\
      \cite{Bernardi_VDF_2010}         & 1.8$\times10^{-3}$ & 166.5 & 2.54  & 2.93   \\
      \hline
      \hline
  \end{tabular*}
  \begin{flushleft}
  \footnotesize \textit{Notes.} The values of $\Phi_\star$ are reported in the adopted cosmology.
  \end{flushleft} 
\end{table}

%######%######%######%######%######%######%######%######%######%######%######%######%######%
\subsection{Stellar VDF vs. Lensing VDF}\label{sec:starVDF_to_lensVDF}

In general, the velocity dispersion $\sigma$ entering the isothermal lens model is different to the stellar velocity dispersion $\sigma_\star$.
The ratio $f_e \equiv \sigma_{e/2}/\sigma$ is used to relate the two quantities, where $\sigma_{e/2}$ is the stellar velocity dispersion measured within half of the effective radius (i.e. a reasonable estimate of $\sigma_\star$).
At $z \leq 0.33$, combined studies of lensing and dynamics for lens galaxies showed that $f_e \approx 1$ on average (\citealt{Kochanek_2000_fe}, \citealt{Treu_2006}, \citealt{Bolton_2008_fe}, \citealt{Grillo_2008_fe}, \citealt{Auger_2010_fe}, \citealt{Zahid_2018_fe}). 
This means that at the angular scales relevant for strong lensing and at low lens redshifts, the average measured central stellar velocity dispersion matches the velocity dispersion for an isothermal mass distribution that reproduces the observed Einstein radius.
This fact has motivated all previous studies in the literature to work under the assumption that the velocity dispersion entering the lensing model $\sigma$ is tracing \textit{exactly} the central stellar velocity dispersion $\sigma_\star$, and directly use the fiducial VDF measured locally from stellar spectroscopy to model the population of lenses.
However, we argue that the spread in the relation between $\sigma$ and $\sigma_\star$ requires treating the VDFs associated to these quantities separately. 
We transform between the measured stellar velocity dispersion function d$\Phi/$d$\sigma_\star$ and the isothermal lens velocity dispersion function d$\Phi/$d$\sigma$ using a convolution with the Gaussian distribution of $f_e \sim \mathcal{N}(\mu_{f_e}; \sigma_{f_e})$ as
\begin{equation}
\dv{\Phi}{\sigma_\star}(\sigma_\star) = 
\int \dd\sigma\dv{\Phi}{\sigma}(\sigma)
\mathcal{N}(\sigma_\star-\mu_{f_e}\sigma; \sigma_{f_e}\sigma)
\end{equation}

To mantain a VDF parametrization of the isothermal VDF, instead of directly deconvolving the measured fiducial stellar VDF we forward model the parameters of the isothermal VDF to match the stellar velocity dispersion profiles at $z=0$, and recover the measured VDF within 5\% in the range $175<\sigma<325$.
We discuss the measurement if $f_e$ in Sect. \ref{sec:Observations}.

\begin{table}
  \caption{Parameters for the $z=0$ isothermal lens VDF obtained from deconvolving the fiducial VDF listed as Reference.}
  \label{tab:IsoLensVDF_param}
  \centering
  \begin{tabular*}{\linewidth}{@{\extracolsep{\fill}} c c c c c}
      \hline 
      Reference & $\Phi_\star$ [Mpc$^{-3}$]& $\sigma_\star$ [km s$^{-1}$] & $\alpha$ & $\beta$\\
      \hline 
      \hline 
      \cite{Sheth_2003_VDF_from_LF}    & 2.08$\times10^{-3}$ & 102.7 & 7.0   & 2.24   \\
      \cite{Mitchell_2005_VDF_from_LF} & 1.46$\times10^{-3}$ & 103.6 & 7.0   & 2.24   \\
      \cite{Choi_2007_SDSS}            & 2.29$\times10^{-3}$ & 181.1 & 2.7   & 3.30   \\
      \cite{Chae_2010}                 & 4.09$\times10^{-3}$ & 235.1 & 0.8   & 4.90   \\
      \cite{Bernardi_VDF_2010}         & 1.35$\times10^{-3}$ & 180.4 & 3.5   & 3.63   \\
      \hline
      \hline
  \end{tabular*}
  \begin{flushleft}
  \footnotesize \textit{Notes.} The values of $\Phi_\star$ are reported in the adopted cosmology.
  \end{flushleft} 
\end{table}

\section{Sample of observed lenses}\label{sec:Observations}
We first introduce the sample of lens galaxies used before discussing our lens model in Section \ref{sec:lens_model}.
To constrain the VDF redshift evolution we fit modeled redshift and Einstein radii distributions from Eq. (\ref{eq:Prob_obs}) to an observed distribution of lenses.
To ensure homogeneous selection criteria, we chose to perform our analysis on the Strong Lensing Legacy Survey lens search (SL2S; \citealt{CFHTLS_SL2S}), limiting the sample to the 25 grade A lenses presented in \cite{Sonnenfeld_SL2S_2013a} that have measured values of the redshift of both the deflector galaxy and the lensed source, and of the Einstein radius and the velocity dispersion of the lens.
The SL2S sample is composed of lenses initially identified in imaging data from the Canada–France–Hawaii Telescope (CFHT) Legacy Survey and then followed up with high resolution imaging from the Hubble Space Telescope and optical/near-infrared spectroscopy.
The 25 lenses in the sample considered in this study have lens and source redshifts in the range $0.238 < z_l < 0.783$ and $1.2 < z_s < 3.48$, respectively.
The Einstein radii span is $0.74^{\prime\prime} < \theta_E < 2.55^{\prime\prime}$.

This SL2S subsample has been modeled combining stellar dynamics and lensing (\citealt{Sonnenfeld_SL2S_2013a}), so each galaxy has a measure of the total mass density slope $\gamma$ and the ratio of measured stellar velocity dispersion and `lensing' velocity dispersion $f_e$.
In this survey, the average values of the recovered total mass slope and the velocity dispersion ratio are $\langle\gamma\rangle\approx 2$ (i.e. isothermal) and $\langle f_e \rangle \approx 1$, respectively.
While there is a strong linear correlation between the values of $\gamma$ and $f_e$ derived combining strong lensing and stellar dynamics, this is due to the way the mass slope is obtained from observables (\citealt{SLACSVIII_Treu}), and the relation might change when using more detailed dynamics models (e.g. including orbits anisotropy, \citealt{ProjectDinos1}).

\begin{figure}
  \includegraphics[width=\linewidth]{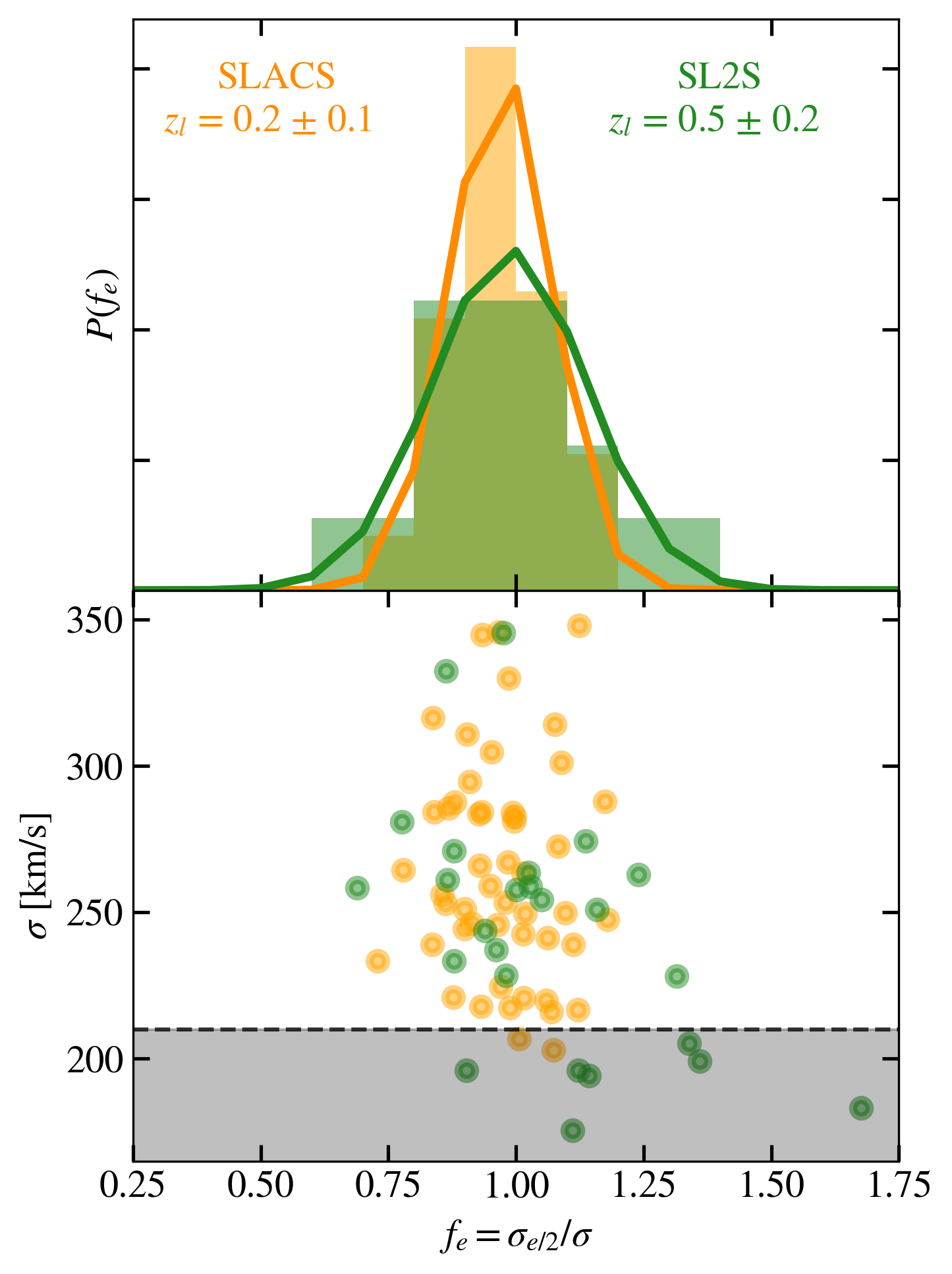}
    \caption{
    Distribution of the ratio of the observed stellar velocity dispersion and the SIE model velocity dispersion, $f_{e} = \sigma_{e/2}/\sigma$ in the SLACS and SL2S samples. 
    The top panel shows the distributions of $f_e$ along with their best-fit Gaussian profile, while the bottom panel shows the distribution of $f_e$ versus $\sigma$.
    The grey shaded are in the bottom panel indicates the lenses with $\sigma<210$ excluded from the Gaussian fit to minimize completeness bias.}
    \label{fig:gamma_fe}
\end{figure}

We fit the distribution of $f_e$ in the SL2S sample with a Gaussian profile and use it to map the measured stellar VDF to the isothermal lens model VDF via a convolution, as discussed in Section \ref{sec:starVDF_to_lensVDF}.
Since the velocity dispersion ratio $f_e$ distribution could change with redshift, we calibrate the fiducial local VDFs by considering the SLACS sample of galaxy scale lenses (\citealt{Bolton_SLACS_2008}, \citealt{Auger_2010_fe}), which has deflectors in the range $0.1 \lesssim z_l \lesssim 0.3$, and has measurements of the ratio $f_e$ obtained with the same method as in the SL2S sample.
At fixed stellar mass, the SLACS lenses have a steeper density profile and larger velocity dispersion than the parent population of galaxies (\citealt{SLACS_debiased}); but are \REPLY{nearly} indistinguishable from non-lens galaxies at fixed velocity dispersion (\citealt{Treu_2006}, \citealt{SLACS_debiased}).
\REPLY{For these reasons, we use the SLACS sample to obtain the local isothermal VDF from the stellar velocity dispersion, but we do not use the SLACS sample in addition to the SL2S sample when performing fits to lens observables.} 
We fit the distributions of values of $f_e$ in SLACS and SL2S with a Gaussian profile, considering only deflectors with $\sigma>210$ km/s, to match the two samples and avoid completeness issues at lower values of $\sigma$, as shown in Fig. \ref{fig:gamma_fe}.
The best fitting Gaussian for $f_e$ (corrected for the intrinsic error on $\sigma_\star$) in the SL2S sample has mean $0.99$ and standard deviation $0.15$, while the we find a mean of $0.97$ and a standard deviation of $0.10$ for $f_e$ in the SLACS sample.
We then use this distribution as a kernel to deconvolve the VDF($\sigma_\star$) to a distribution function in $\sigma$ which can be compared directly with the VDF from lensing at $z=1$. We discuss the predicted lens properties given this VDF in the next section. 
This step is shown in the transition from the lower to the upper panel on the left hand side of Fig. \ref{fig:VDF_results_nu}.
The resulting best-fitting VDF parameters for the isothermal lens VDF at $z=0$ for each of the fiducial local stellar VDFs are listed in Table \ref{tab:IsoLensVDF_param}.

% The redshifts in SL2S have been obtained from ground-based spectroscopy, and we model selection based on redshift determined based on only the visible source redshift intervals as described in Appendix \ref{sec:Appendix}.

%######%######%######%######%######%######%######%######%######%######%######%######%######%

\section{Lensing model}\label{sec:lens_model}

% We constrain the VDF evolution from a sample of observed galaxy scale strong lenses, using the distributions of lens and source redshifts and Einstein radii.
% Compared to the well tested lens-redshift test (\citealt{Kochanek_1992}), which is based on the differential multiple image optical depth as a function of the redshift and size of the lens population, our modeling approach is limited to samples of galaxy-galaxy lenses but has the advantage of explicitly modeling the effect of the survey and lens identifiability constraints on the observed redshifts and size distributions (\citealt{Ferrami_Wyithe_Lens_stat_model}).

In this paper we focus on the VDF of early-type galaxies (ETGs), and adopt the usual approximation that the deflector population is dominated by ETGs (e.g., \citealt{Turner_1984}, \citealt{Kochanek_1996}, \citealt{Oguri_Marshall}).
Our lens model is based on a Singular Isothermal Ellipsoid (SIE) mass distribution, which approximates the total matter surface density profile within a typical strong lensing scale of $\sim 1$ arcsec for $z \lesssim 1$.
(\citealt{Gavazzi_SLACS_2007}, \citealt{Koopmans_bulge_halo_conspiracy}, \citealt{Lapi_2012}, \citealt{Sonnenfeld_SL2S_2013}, \citealt{SLACS_debiased}).
After marginalizing over source position, the strong lensing effects of a SIE lens depend uniquely on the total lens mass, defined by the \textit{isothermal} velocity dispersion $\sigma$, and on the ellipticity of the SIE mass distribution.
The ellipticity $\varepsilon$ of the SIE mass distribution affects the total area within the outer caustic by a factor $C(\varepsilon) \lesssim 1$. 
Averaging over a fiducial early-type galaxies ellipticity distribution (\citealt{van_der_Wel_2014}), we find $\overline{C} \approx 0.9$ (\citealt{Ferrami_Lensing_bright_end}). The ellipticity distribution of lenses is considered constant with $z$.
For a SIE lens model the Einstein radius is a function of the velocity dispersion of the deflector $\sigma$ and of the angular diameter distance\footnote{The cosmological distances $D_s$, $D_{ls}$ are the angular diameter distances between the source and the observer or deflector, respectively.} $D$ as
\begin{equation}\label{eq:Einstein_radius}
  \theta_E(\sigma, z_l, z_s) = 4\pi\frac{D_{ls}}{D_s}\left(\frac{\sigma}{c}\right)^2 = 1^{\prime\prime} \frac{D_{ls}}{D_s}\left(\frac{\sigma}{186 \text{ km/s}}\right)^2 \text{ .}
\end{equation}
If the source and lens redshifts are known, the Einstein radius therefore gives a measure of $\sigma$ for a fixed cosmology.

We follow the model described in \cite{Ferrami_Wyithe_Lens_stat_model} to construct the expected distributions of lens and source redshifts and Einstein radii of a sample of \textit{identifiable} lenses.
These distributions are a function of the assumed cosmology, of the lens mass function evolution, of the survey parameters (e.g., seeing, depth), and of the features that are used by a given lens search method to identify the candidates.
For a lens with fixed velocity dispersion $\sigma$ and redshift $z_l$, the number density of background sources with absolute magnitude $M$ at redshift $z_s$ that are lensed is obtained as the product between the luminosity function and the comoving volume behind this single lens (calculated using the area enclosed in the portion of the source plane that can produce multiple images).
In a flat universe, this equates to
\begin{equation}\label{eq:Prob_lens_backgnd}
  \frac{\dd N_{SL}(M, z_s | \sigma, z_l)}{\dd z_s \dd M}= C\pi  \theta_E^2 D_c^2(z_s)\frac{c}{H(z_s)} \Psi_B(M, z_s)\:,
\end{equation}
where $D_c$ is the comoving distance to the source and $\Psi_B$ is the biased luminosity function.
Accounting for the lens identifiability constraints $I$ imposed by the survey design and the lens detection method adopted, we can write 
\begin{equation}\label{eq:Prob_lens_backgnd_with_fractions12}
\begin{aligned}
\frac{\dd N_\text{SL}^\text{obs}}{\dd z_s \dd M}(M, z_s, I) =
\frac{\dd N_\text{SL}}{\dd z_s \dd M} 
\times \max(\mathcal{S}_1\mathcal{F}_\text{arc}, \mathcal{S}_2\mathcal{F}_\text{2nd}) \mathcal{S}_E \:,
\end{aligned}
\end{equation}
where $\mathcal{S}_{1/2}$, $\mathcal{F}_{1/2}$, and $\mathcal{S}_E$ are step functions depending on the detection signal-to-noise ratio, on the fraction of bright images and on the PSF/seeing limited image resolution, respectively.

Assuming a survey with solid angle $A_s$, the total number of lensed sources in a given field per unit source and lens redshift is
\begin{equation}\label{eq:Number_lens_galaxies_redsh_sigma}
\begin{aligned}
  \frac{\dd N_\text{SL}^\text{obs}}{\dd z_l \dd z_s \dd \sigma} =
  \int_0^{M_\text{cut}(z_s)} \dd{M} 
  \dv{V(A_s, z_l)}{z_l} \frac{\dd N_\text{SL}^\text{obs}}{\dd z_s \dd M}
  \Phi(\sigma, z_l) \:,
\end{aligned}
\end{equation}
where $M_\text{cut}(z_s)$ is the magnitude cut, and $\dv{V}{z_l}$ is the comoving volume shell within $A_s$ multiplied by the VDF $\Phi(\sigma, z_l)$.

We obtain the probability distribution of finding a lens with the set of observables $[z_l, z_s, \theta_E]$ as
\begin{equation}\label{eq:Prob_obs}
  P(z_l,z_s, \theta_E) = \frac{1}{\mathcal{N}} \frac{\dd N_\text{SL}^\text{obs}}{\dd z_l \dd z_s \dd \sigma} \dv{\sigma}{\theta_E} \:,
\end{equation}
where $\mathcal{N}$ is the appropriate normalization term.
By comparing an observed sample of lenses with this probability distribution rather then the lens system number density over the same parameter space, we are not sensitive to the parameters entering the lens distribution model as  multiplying constants, such as $\Phi_{\star,0}$, the survey area and completeness\footnote{We assume that the completeness is not redshift dependent.}, the Hubble radius $c/H_0$, or the correction factor to the caustic area introduced by lens ellipticity.
The normalization factor is weakly degenerate with the ellipticity distribution via the caustic area correction factor $C$, but we ignore this contribution to its evolution.
On the other hand, this choice induces a partial decrease in sensitivity to the VDF number density evolution parameter (i.e., $\nu_n$ or $P$), as noted for the lens-redshift test (\citealt{Ofek_2003}).

%######%######%######%######%######%######%######%######%######%######%######%######%######%

\section{Analysis and results}\label{sec:Analysis}

In this section we use lens observables from SL2S to constrain the VDF evolution.
The redshift evolution of the modeled lens distribution in Eq. (\ref{eq:Number_lens_galaxies_redsh_sigma}) depends to the evolution of the source population, the lens detection method requirements and on the deflector population mass distribution expressed as the VDF.
We assume the source population evolution from the UV luminosity function described in \cite{Bouwens21_data}, and impose the lens detection parameters for the SL2S survey as described in \cite{Ferrami_Wyithe_Lens_stat_model}.
In the underlying lensing statistics model we also assume the luminosity-size relation taken from \cite{Shibuya_L_size_rel} and calculate the lens light profile through a linear fit of the Fundamental Plane parameters between $z=0$ and $z=2$ (\citealt{Stockmann_2021}).
The resulting model probability distribution is solely a function of the VDF profile, and in particular of the redshift evolution $\Phi_{\rm{s}}$ and $\sigma_{\rm{s}}$ according to the scaling relations in Eqs. (\ref{eq:scaling_power_law}) or (\ref{eq:scaling_exp}).

%######%######%######%######%######%######%######%######%######%######%######%######%######%

\subsection{Maximum likelihood}\label{sec:Likelihood}

Given a set of $n$ lenses with observed parameters $[\bar z_l, \bar z_s, \bar \theta_E]$, we fit each scaling relation individually with the \textsc{emcee} Markov Chain Monte Carlo (MCMC) sampler developed by \cite{emcee} on the log likelihood estimator $\log\mathcal{L}$ defined as
\begin{equation}\label{eq:Likelihood}
\ln\mathcal{L}(\{X\}) = \sum_{i=1}^n \ln{P_i\left(\bar z_{l(i)}, \bar z_{s(i)}, \bar \theta_{E(i)} \biggr| \{X\}\right)},
\end{equation}
where $\{X\}$ is either $\{\nu_n, \nu_v\}$ or $\{P, U\}$ (Eqs. \ref{eq:scaling_power_law} - \ref{eq:scaling_exp}).

We run each MCMC until the the autocorrelation time $\tau_\text{MC}$ estimation converges within one percent and the chain length exceeds $100 \tau_\text{MC}$.
To avoid biases induced by the choice of initial conditions of the walkers and ensure that each step is uncorrelated, we then discard the first $2 \tau_\text{MC}$ steps as the burn-in phase of each chain and thin each chain by subsampling it every $0.5 \tau_\text{MC}$.
%######%######%######%######%######%######%######%######%######%######%######%######%######%
\subsection{Constraints on VDF evolution}

\begin{figure*}
  \includegraphics[width= 0.8\linewidth]{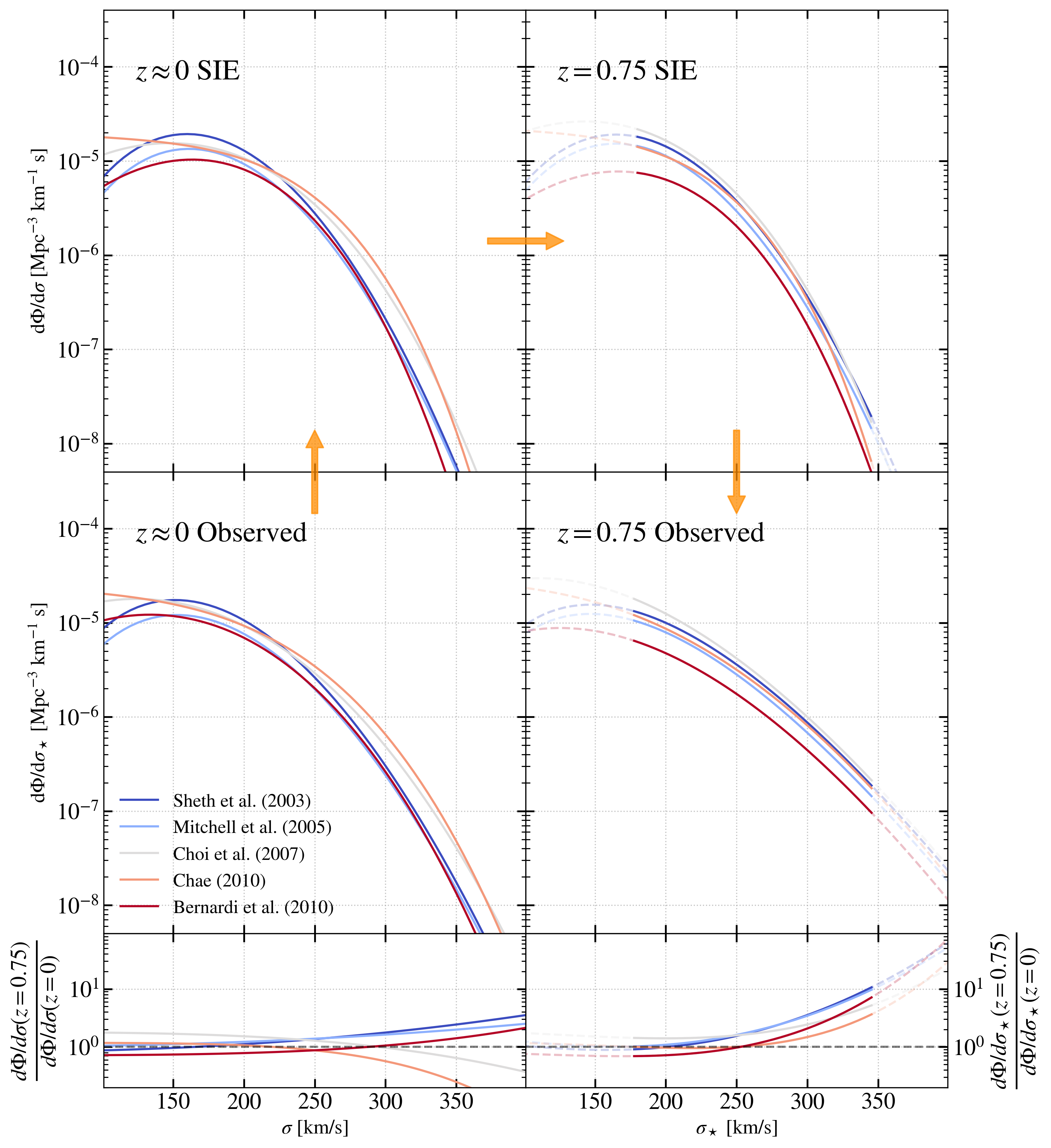}
    \caption{
    The steps to derive the redshift evolution of the stellar VDF, assuming a power law scaling relation. 
    The lower left panel shows the fiducial stellar VDF measured in the local universe, listed in Table \ref{tab:Fiducial_param}.  
    The colors in each panel match the associated redshift $z=0$ assumed stellar VDF parametrization.
    The upper left panel shows the local isothermal lens VDF, obtained deconvolving the stellar VDF with the distribution of $f_e = \sigma_{e/2}/\sigma$. 
    The upper right plot shows the evolution of the isothermal lens VDF at $z=0.75$, obtained via an MCMC fitting agains the SL2S (\citealt{Sonnenfeld_SL2S_2013a}) sample of strong lenses.
    The lower right plot shows the $z=0.75$ stellar VDF, obtained convolving the isothermal lens VDF with the $f_e$ distribution.
    The extension plots at the bottom right shows the ratio between the $z=0.75$ and the local isothermal (left) or stellar (right) VDFs.
    \REPLY{The solid lines in the the panels on the right indicate the range of stellar velocity dispersion spanned by the SL2S data used to constrain the VDF redshift evolution.}}
    \label{fig:VDF_results_nu}
\end{figure*}

We next present the evolution of the VDF normalization and scale-velocity with cosmic time.
We start with the local isothermal lens VDF (upper left panel in Figure \ref{fig:VDF_results_nu} and Table \ref{tab:IsoLensVDF_param}) as described in Sec. \ref{sec:starVDF_to_lensVDF}.
We perform the MCMC fitting on the VDF evolution parameters and obtain an estimate of the evolved isothermal VDF at $z=0.75$ (upper right panel).
We then convolve this with the Gaussian distribution fitted on the $f_e$ values from the SL2S distribution ($z\approx0.5$) to retrieve the evolved stellar VDF (lower right panel).
We perform this analysis on all five fiducial stellar VDFs, each time considering both scaling relations in Eqs. (\ref{eq:scaling_power_law}) or (\ref{eq:scaling_exp}).

\begin{figure}
  \includegraphics[width=\linewidth]{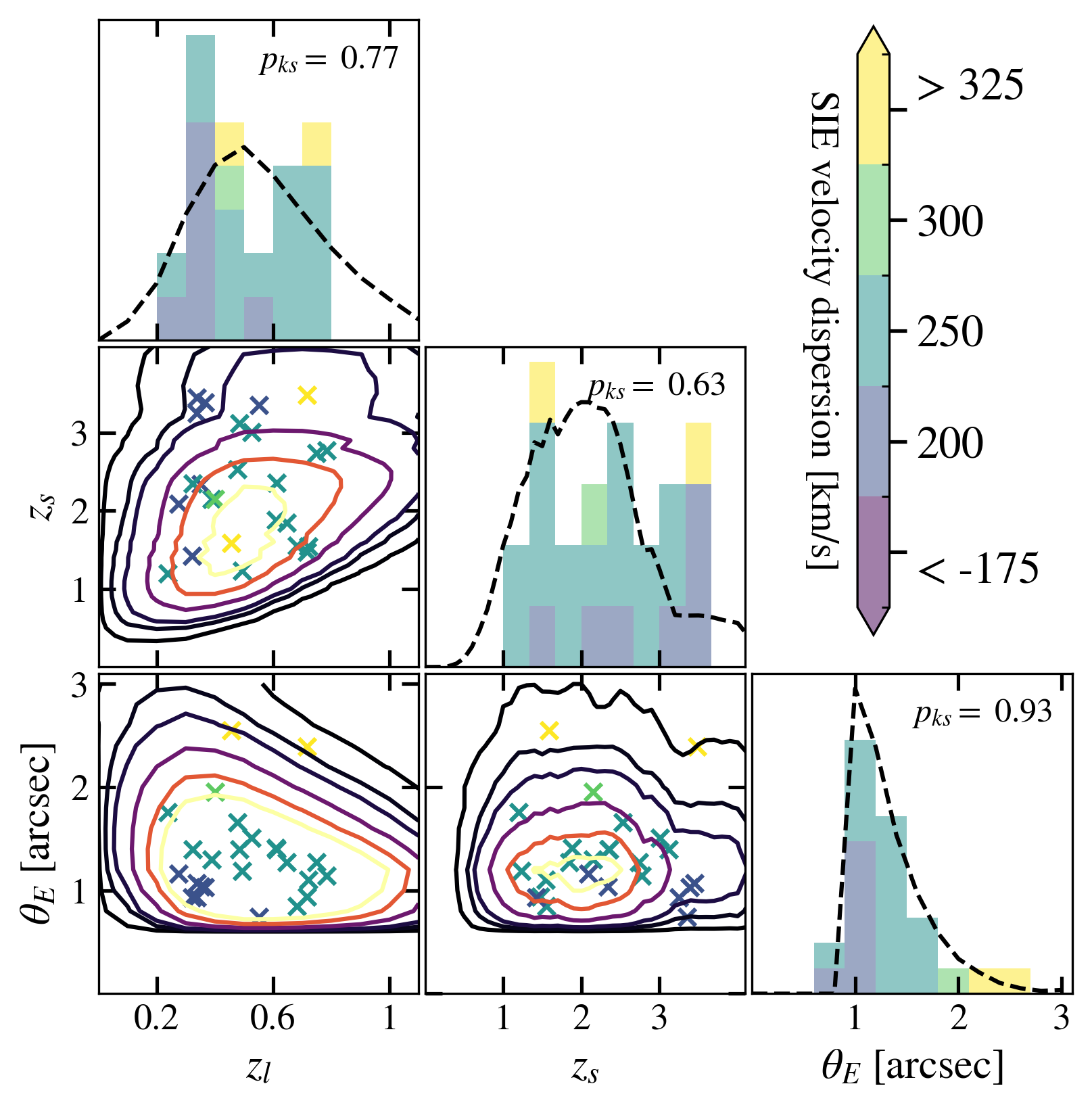}
    \caption{Distribution of the sample of observed grade A lenses in SL2S (\citealt{Sonnenfeld_SL2S_2013a}) against the model predictions associated to the best fit VDF evolution assuming the local distribution from \citealt{Choi_2007_SDSS}.
    The figure shows the 2D projections and marginalized probability distributions of the three dimensional parameter space composed by lens and source redshifts and the size of the Einstein radius.
    The contour plot shows the model iso-probability contours (solid lines) against the SL2S sample (crosses and stacked histograms). The observed sample is color coded based on their SIE velocity dispersion obtained inverting Eq. (\ref{eq:Einstein_radius}).
    The p-value associated to the Kolmogorov-Smirnov test between the marginalized probability distribution of the model and the normalized histogram of the observed sample is reported on the top right corner of the plots on the diagonal.}
    \label{fig:contours_distr}
\end{figure}

As an example, the best-fit redshift and Einstein radii distributions are shown in Fig. \ref{fig:contours_distr} alongside the distributions of the observed SL2S sample.
The panels on the diagonal present the average p-value associated to the Kolmogorov-Smirnov test between the observed lens distributions and one hundred realizations of 25 galaxies drawn from the best-fitting modeled 1-d probability distributions, indicating that the modeled redshifts and size distributions are compatible with the data.
The excess of $z_s\gtrsim3$ sources and $z_l\gtrsim0.6$ lenses are statistical fluctuation due to the small sample size, but are not significant ($p_{KS} = 0.63$ and $p_{KS} = 0.77$, respectively).
If real these might be the effect of redshift-dependent completeness induced by the lens search method.

\begin{figure}
  \includegraphics[width=\linewidth]{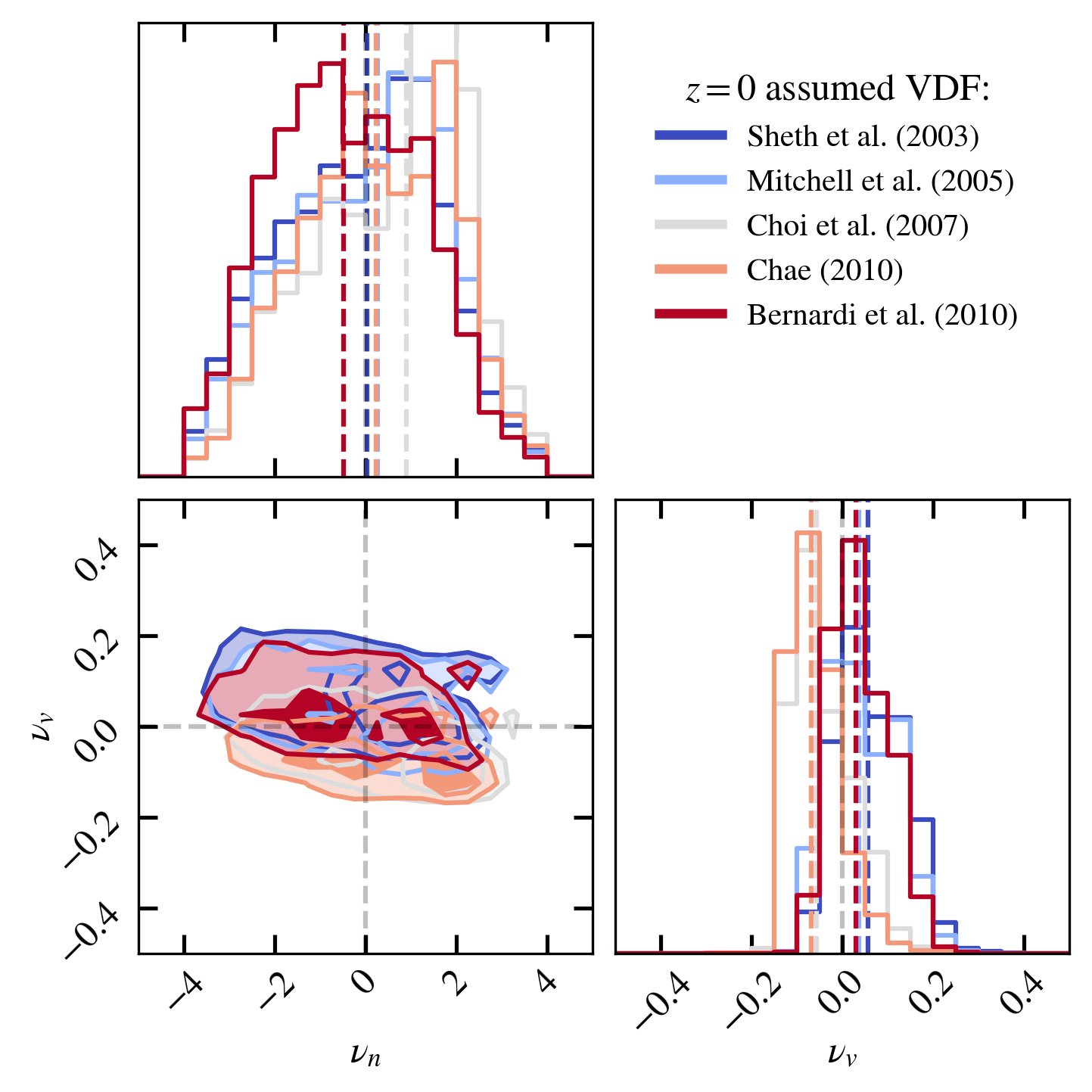}
    \caption{
    Constraints on the redshift evolution of the VDF assuming a power-law scaling relation (Eq.\ref{eq:scaling_power_law}). 
    The derived $\nu_n$-$\nu_v$ 1 and 2-sigma contours for each choice of the redshift $z=0$ VDF is shown with the same color as in the left panel of Figure \ref{fig:VDF_results_nu}.}
    \label{fig:Chains_nu}
\end{figure}

\begin{figure}
  \includegraphics[width=\linewidth]{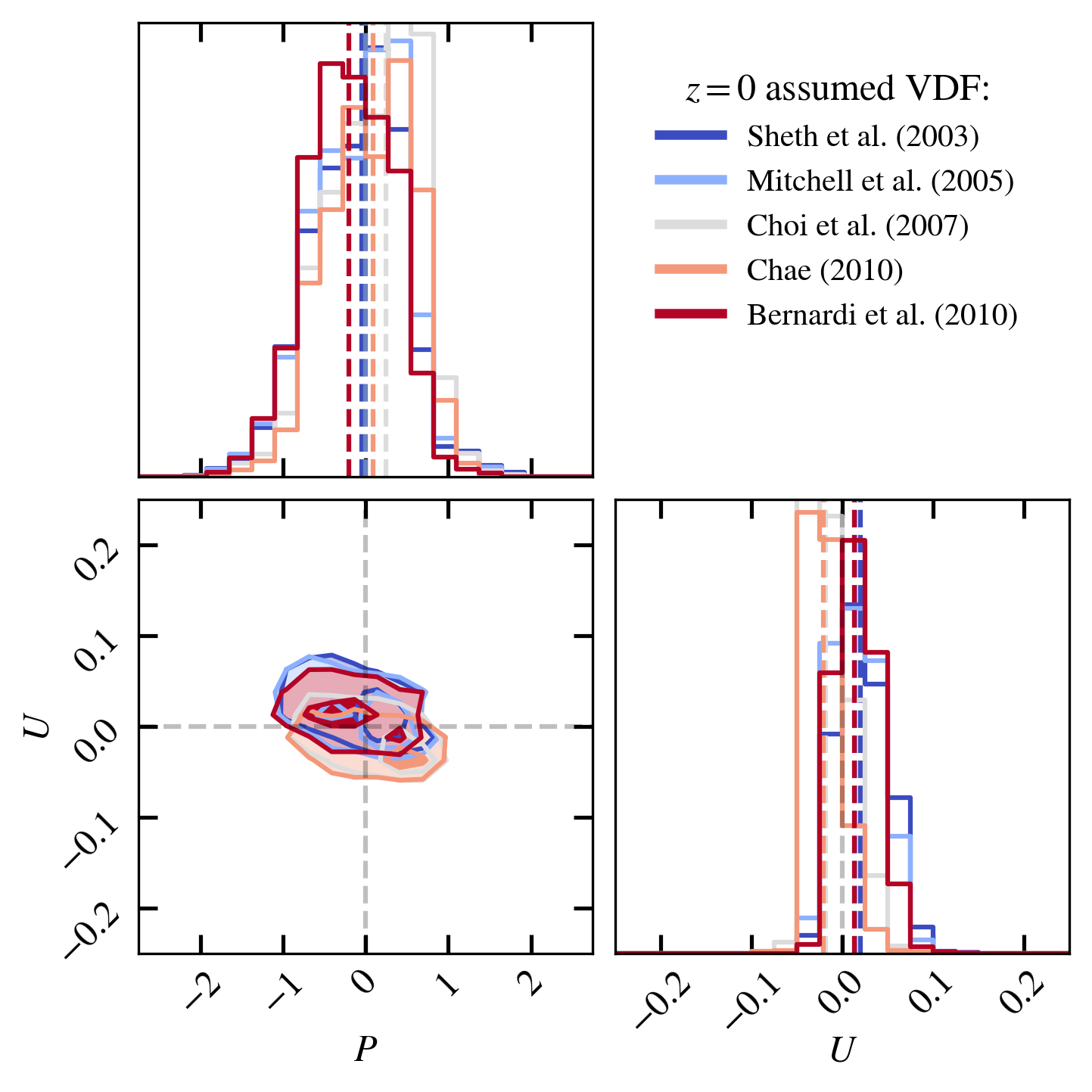}
    \caption{
    Constraints on the redshift evolution of the VDF assuming an exponential scaling relation (Eq.\ref{eq:scaling_exp}).  
    The derived $P$-$U$ 1 and 2-sigma contours for each choice of the redshift $z=0$ VDF is shown with the same color as in the left panel of Figure \ref{fig:VDF_results_nu}.}
    \label{fig:Chains_PU}
\end{figure}

Figure \ref{fig:VDF_results_nu} shows the resulting profiles for each MCMC chain on the VDF evolution parameters, evaluated at $z=0.75$ in the upper right panel.
We see that the best fitting results for each pair of local VDF and scaling relation converges to a similar VDF profile at large values of $\sigma$.
This suggests that the method is consistently able to find the VDF evolution parameters that lead to the best fit mass density evolution of the deflector population, independent of the assumptions on the redshift $z=0$ population. In particular, the model is able to constrain the high end of the VDF for $\sigma\gtrsim 250$ ($\approx 1''$ for the average lens and source redshifts of the SL2S sample in Eq. \ref{eq:Einstein_radius}), corresponding to the most effective deflectors.
We stress that this convergence is in spite of the fact that our constraints do not include lens numbers\footnote{Lens numbers, which provide an estimate of the lensing optical depth, have large systematic uncertainties \cite{Barone_Nugent_2015}.}.
One can see that the main effect of the convolution to map $\dv{\Phi}{\sigma}$ to $\dv{\Phi}{\sigma_\star}$ is to increase the density of high-$\sigma_\star$ galaxies ($\sigma_\star \gtrsim 250$ km/s).
Looking at the ratio between the $z=0.75$ and the local VDFs at the bottom of Fig. \ref{fig:VDF_results_nu} for the isothermal (left subplot) and stellar velocity dispersion (right subplot), it appears that the observed over-abundance of $\sigma_\star \gtrsim 200$ km/s galaxies is due to the increased scatter in the $\sigma$-$\sigma_\star$ relation rather than the evolution of the underlying isothermal VDF.\\
\indent Our constraints on the VDF evolution are listed in Table \ref{tab:VDF_z_evo}, and the MCMC chains for the power-law scaling relation and the exponential scaling relation are plotted in Figs. \ref{fig:Chains_nu} and \ref{fig:Chains_PU}, respectively.
The evolution of the scaling parameters can be divided in two groups: the one having the local VDF from \cite{Sheth_2003_VDF_from_LF}, \cite{Mitchell_2005_VDF_from_LF}, and \cite{Bernardi_VDF_2010} that show a slow negative evolution in normalization and no evolution in scale velocity (e.g. $\nu_n \approx -0.5$ and $\nu_v \approx 0$), and the second group having the local VDF from \cite{Choi_2007_SDSS} and \cite{Chae_2010} showing a steeper positive normalization evolution and a mild negative scale velocity evolution (e.g. $\nu_n \approx 1.2$ and $\nu_v \approx -0.12$). 
We note that the first group is also compatible with no evolution within 1-$\sigma$, while the second group only within 2-$\sigma$.
Matching trends can be seen also in the set of results associated with an exponential scaling relation.
Comparing the upper left panel to the upper right panel of Figure \ref{fig:VDF_results_nu}, this can be understood by the fitting procedure trying to match the the high-$\sigma$ end of the VDF of the two groups, driving opposite signs in the evolution parameters.\\

\begingroup
\begin{table*}
  \parbox{13.5cm}{\caption{Summary of the constraints on the redshift evolution of the VDF parameters.
  Median values of the likelihood function in the $\nu_n$-$\nu_v$ and $P$-$U$ planes, for the SL2S sample of grade A lenses. The first column, indicates the assumed profile of the VDF at $z=0$ .}
  \label{tab:VDF_z_evo}}
  \centering
  \renewcommand{\arraystretch}{1.34}
  \begin{tabular*}{0.76\linewidth}{@{\extracolsep{\fill}} c c c c c}
      \hline 
      Reference & 
      $\nu_n=$ d$\Phi_\star(z)/$d$z$& 
      $\nu_v=$ d$\sigma_\star(z)/$d$z$ & 
      $P=$ d$\log_{10}\Phi_\star(z)/$d$z$ & 
      $U=$ d$\log_{10}\sigma_\star(z)/$d$z$\\
      \hline 
      \hline 
      \cite{Sheth_2003_VDF_from_LF} & $-0.686_{-1.410}^{+1.785}$ & $0.042_{-0.054}^{+0.081}$ & $-0.218_{-0.440}^{+0.494}$ & $0.014_{-0.018}^{+0.028}$ \\
      \cite{Mitchell_2005_VDF_from_LF} & $-0.326_{-1.629}^{+1.770}$ & $0.020_{-0.054}^{+0.083}$ & $-0.166_{-0.457}^{+0.528}$ & $0.009_{-0.020}^{+0.027}$ \\
      \cite{Choi_2007_SDSS} & $1.413_{-2.221}^{+1.093}$ & $-0.112_{-0.047}^{+0.093}$ & $0.413_{-0.667}^{+0.343}$ & $-0.038_{-0.016}^{+0.031}$ \\
      \cite{Chae_2010} & $0.984_{-1.917}^{+1.370}$ & $-0.127_{-0.042}^{+0.055}$ & $0.281_{-0.592}^{+0.434}$ & $-0.043_{-0.014}^{+0.019}$ \\
      \cite{Bernardi_VDF_2010} & $-0.490_{-1.552}^{+1.788}$ & $0.001_{-0.046}^{+0.076}$ & $-0.151_{-0.466}^{+0.528}$ & $0.002_{-0.017}^{+0.026}$ \\

      \hline
      \hline
  \end{tabular*}
\end{table*}
\endgroup

%######%######%######%######%######%######%######%######%######%######%######%######%######%
\subsection{Comparison with previous studies}\label{sect:Comparison}

\begin{figure}
  \includegraphics[width=\linewidth]{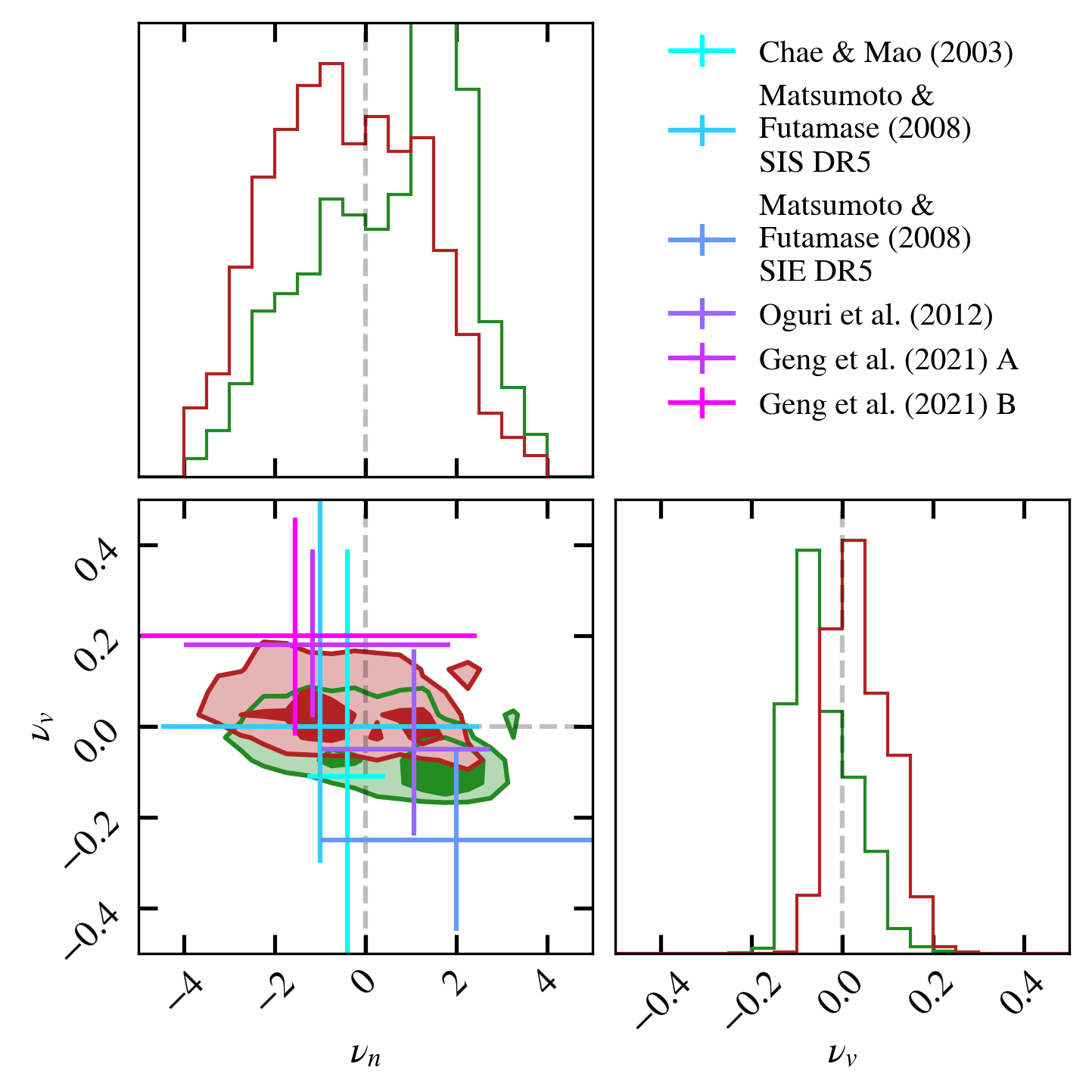}
    \caption{Constraints on the redshift evolution of the VDF given by our model assuming a power law scaling relation and a local VDF as described in \citet{Choi_2007_SDSS} or \citet{Bernardi_VDF_2010} (green and red 1 and 2-sigma contours, respectively), compared with a sample of past studies that used strong lensing to constrain the VDF.
    }
    \label{fig:param_comp_nu}
\end{figure}

\begin{figure}
  \includegraphics[width=\linewidth]{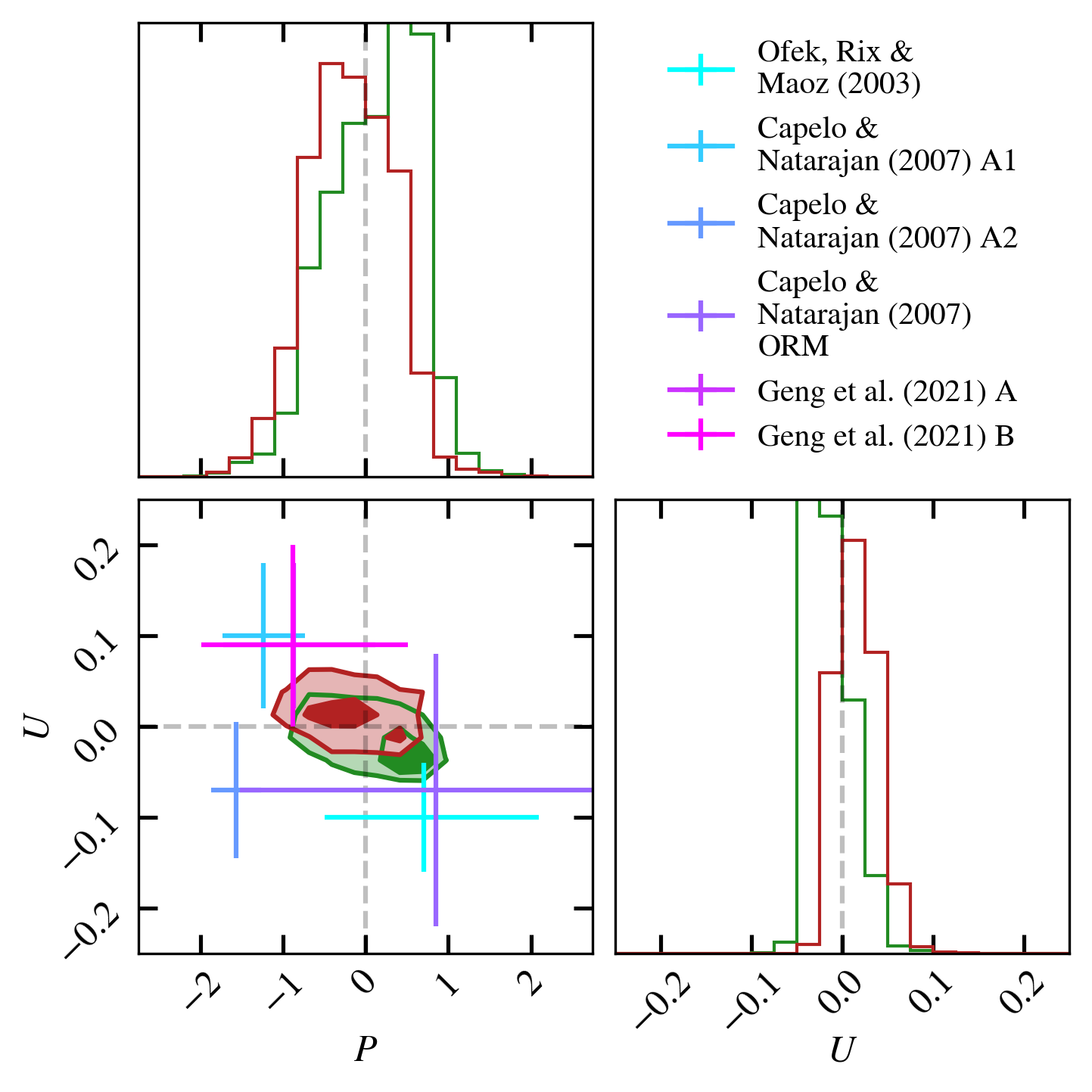}
    \caption{Constraints on the redshift evolution of the VDF given by our model assuming an exponential scaling relation and a local VDF as described in \citet{Choi_2007_SDSS} or \citet{Bernardi_VDF_2010} (green and red 1 and 2-sigma contours, respectively), compared with a sample of past studies based on the lens-redshift test.}
    \label{fig:param_comp_PU}
\end{figure}

We compare our constraints on VDF evolution to a set of previous studies that give constraints on the power-law and exponential evolution models with gravitational lensing (Figures \ref{fig:param_comp_nu} and \ref{fig:param_comp_PU}).
In Figure \ref{fig:param_comp_nu}, we show our results on the power-law scaling evolution for the local VDF parametrization of \cite{Choi_2007_SDSS} and \cite{Bernardi_VDF_2010}, and compare with the results of a sample of past studies.
These past studies include the test on the observed image separations and rate of multiple imaging of 13 lenses from \cite{ChaeMao2003_VDFevo}, the lens-redshift test performed by \cite{Matsumoto_Futamase_2008} on 13 lenses, the analysis on redshift and angular size distribution performed on a sample of 19 quasars in \cite{OguriVDF2012}, and the lens-redshift test presented in \cite{Geng_2021} based on two samples of 158 (sample A) and 126 (sample B) strong lensing systems obtained through different search methods from the SLACS, BELLS, SL2S and LSD surveys.
\footnote{The cosmology assumed in \cite{ChaeMao2003_VDFevo} is flat $\Lambda$CDM with $\Omega_m = 0.3$, while \cite{Matsumoto_Futamase_2008} and \cite{OguriVDF2012} use the cosmology parameters from \textit{WMAP} and \cite{Geng_2021} uses the cosmological parameters from \textit{Planck}.}

Similarly, in Figure \ref{fig:param_comp_PU} we plot our results on the exponential scaling evolution  based on the local VDF parametrization of \cite{Choi_2007_SDSS} and \cite{Bernardi_VDF_2010} compared with the lens-redshift test from \cite{Ofek_2003} and in \cite{Capelo_Natarajan_2007} and \cite{Geng_2021}.
The sample of \cite{Ofek_2003} consists of 71 lenses from the CASTLES sample, while \cite{Capelo_Natarajan_2007} uses three samples of 42 (A1), 70 (A2) lenses each, and the sample from \cite{Ofek_2003}. 
\footnote{The cosmology assumed in \cite{Ofek_2003} and \cite{Capelo_Natarajan_2007} is flat $\Lambda$CDM with $\Omega_m = 0.3$.}

\begin{figure}
  \includegraphics[width=\linewidth]{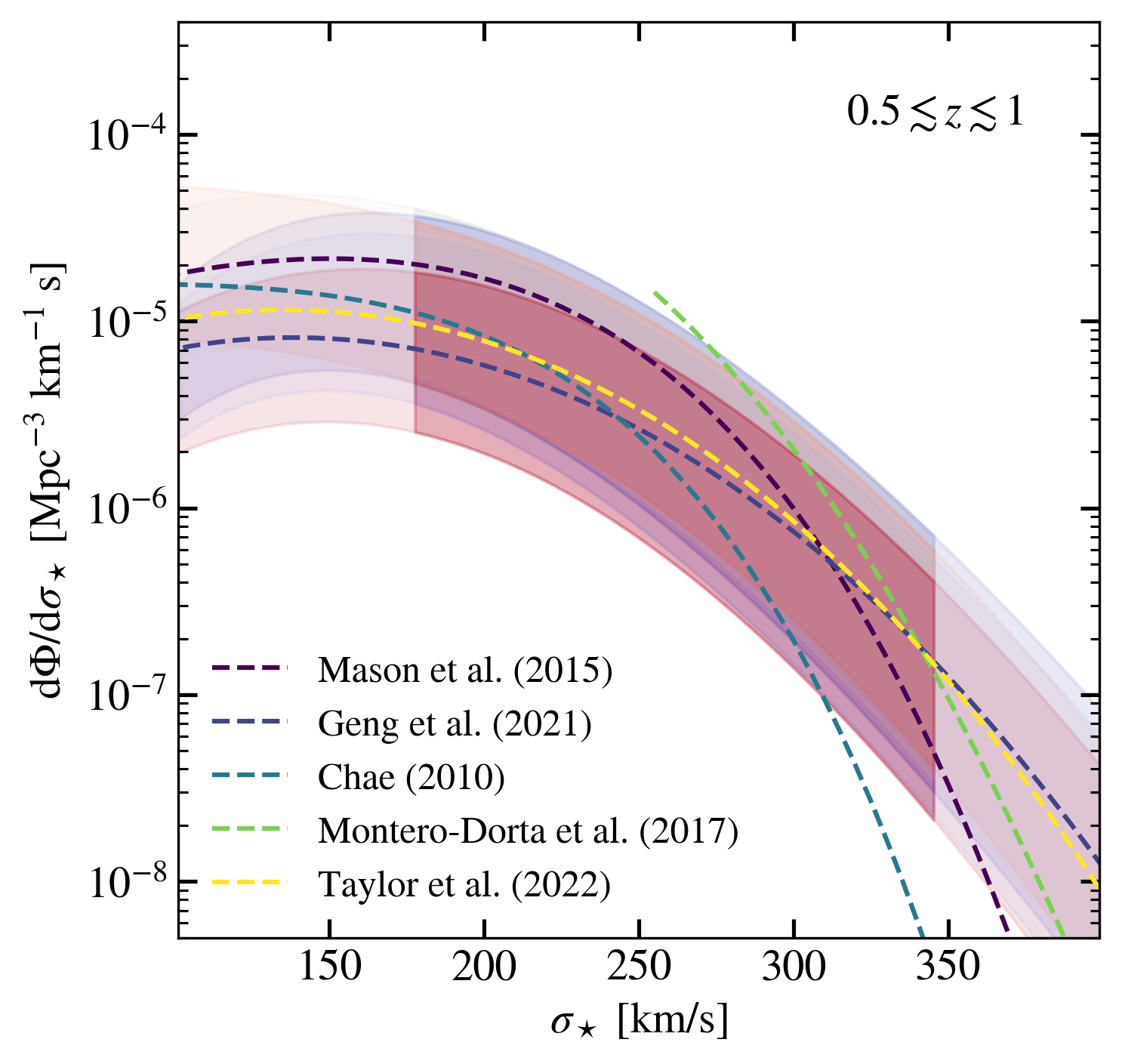}
    \caption{Comparison with measured and inferred VDFs at $0.5\lesssim z \lesssim 1$. 
    The plot shows the 1-$\sigma$ confidence interval for the $z = 0.75$ VDF obtained assuming a power-law redshift evolution, along with a set of observed and inferred VDF at $0.5\lesssim z \lesssim 1$.
    \REPLY{The 1-$\sigma$ bands follow the same color convention introduced in Fig. \ref{fig:VDF_results_nu} for the five local VDF. 
    The range of velocity dispersion values measured in the lens sample is indicated by the darker shaded regions.}}
    \label{fig:VDF_comparison_nu}
\end{figure}

One can see that our model provides constraints on the evolution parameters which are tighter than previous techniques such as optical depth to lensing (ODTL) and the lens-redshift test.
Our constraints also show reduced systematic biases introduced by the lens search process (especially for the exponential scaling relation). 
We also find that while the evolution of the VDF profile converges to similar values for different choices of the fiducial VDF at $z=0$, the values of the VDF evolution parameters may change greatly as a function of the assumed VDF at $z=0$.\\
\indent Figure \ref{fig:VDF_comparison_nu} shows the 1-$\sigma$ confidence intervals of our best fit models based on a power-law scaling to a set of VDFs, measured (\citealt{Montero_Dorta_VDF}, \citealt{Taylor_VDF}) or inferred (\citealt{Chae_2010}, \citealt{Mason_2015}, \citealt{Geng_2021}) at redshift $0.5 \lesssim z \lesssim 1$.
\REPLY{While the redshift distribution of the SL2S deflectors sample ranges between $0.25 \lesssim z_l \lesssim 0.75$ with a peak at $z_l \approx 0.5$, as shown in Figure \ref{fig:contours_distr}, we choose to represent the VDF profile at $z=0.75$ in order to compare with the other published VDFs.}
As introduced in the previous sections, our predicted VDF profiles are convolved with a Gaussian profile in $f_e$ to compare them with the observed stellar central velocity dispersion function.
Our results are compatible with the spectroscopically measured VDFs (\citealt{Montero_Dorta_VDF}, \citealt{Taylor_VDF}) at larger redshift, especially in the high-$\sigma$ regime.

\subsection{Prospects for upcoming large sample of lenses}
By the end of this decade, the sample of lenses with available spectra will increase to a total of $\sim 10^4$ combining Euclid (\citealt{Euclid_Wide_Survey}), Vera Rubin Observatory (\citealt{Vera_Rubin_LSST}), and Roman Space Telescope (\citealt{Roman_WFIRST}) photometrically selected candidates with ground based spectroscopic followup from the 4MOST Strong Lensing Spectroscopic Legacy Survey (4SLSLS; \citealt{4SLSLS_COLLETT}). 
This increase sample size will allow much more stringent tests on the evolution of the deflectors mass function to be performed. 
\REPLY{In addition to the significant increase in the number of homogeneously identified lenses, these upcoming surveys will expand the range of lens properties in the samples to include lower velocity dispersion (i.e. smaller Einstein radii, resolved by space based telescopes) and higher deflector redshifts. For example, around $\sim 15\%$ of lenses discovered in Euclid and Roman will have $150 < \sigma/[\rm{km}\:\rm{s}^{-1}] < 200$ and $\approx8\%$ of lenses above $z_l>1$ (these fractions are obtained by integrating the relevant probability distributions presented in \citealt{Ferrami_Wyithe_Lens_stat_model}). The increase in number of lenses will be important for discriminating between more detailed lens statistics models, and will also allow estimates of a non-parametric form of the VDF profile. In addition, access to new regions of $\sigma-z_l$ parameter space associated with less massive and more distant galaxies will better trace aspects of galaxy evolution.}
Furthermore, the few tens of lenses with $z_l>1$ that are expected to be discovered with James Webb Space Telescope (JWST) within $\approx 1 $deg$^2$ of combined pointings (\citealt{Holloway_2023}, \citealt{Ferrami_Wyithe_Lens_stat_model}), will provide novel constraints on the evolution of the VDF at redshifts above unity.

%######%######%######%######%######%######%######%######%######%######%######%######%######%

\section{Conclusions}

We investigate the use of a strong lensing statistics model to constrain VDF evolution from an observed sample of galaxy-galaxy strong lenses (SL2S, \citealt{Sonnenfeld_SL2S_2013a}).
We focus on the evolution of the normalization and the characteristic velocity of the parametrized VDF (Eq. \ref{eq:VDF}), considering scaling relations that are power law and exponentially evolving in $z$.
Moreover, we assess the robustness of our constraints against different choices of the fiducial VDF in the local universe.
We also account for the spread in the relation between the central stellar velocity dispersion and the `lensing' velocity dispersion associated with isothermal models.
Our results favor a slow evolution of the normalization factor and the characteristic velocity with redshift which is compatible with a no-evolution scenario.
The sign of the redshift evolution parameters depend on the choice of the fiducial profile of the stellar VDF in the local universe. 
Our best-fit evolution parameters yield a VDF at $z\sim0.75$ that is consistent with direct observations, especially for massive $\sigma\gtrsim 250$ galaxies.
This reconciles the tension between directly observed and inferred $z\sim1$ VDFs raised in \cite{Taylor_VDF}, because the over abundance of observed $\sigma_\star \gtrsim 200$ km/s galaxies is due to the scatter in the $\sigma$-$\sigma_\star$ relation rather than intrinsic evolution of the isothermal VDF, which is the quantity constrained by the studies based on strong lensing statistics.

%######%######%######%######%######%######%######%######%######%######%######%######%######%

\section*{Acknowledgments}
\REPLY{We thank the anonymous referee for useful comments that improved the presentation.}
This research was supported by the Australian Research Council Centre of Excellence for All Sky Astrophysics in 3 Dimensions (ASTRO 3D), through project number CE170100013.
 
\section*{Data Availability}

The package used to model the lens statistics, \textsc{galess}, is publicly available\footnote{\url{https://github.com/Ferr013/GALESS}}.
The results of the analysis conducted in this paper will be shared upon reasonable request to the corresponding authors.

\bibliographystyle{mnras}
\bibliography{VDF_MCMC}

\begin{thebibliography}{}
\makeatletter
\relax
\def\mn@urlcharsother{\let\do\@makeother \do\$\do\&\do\#\do\^\do\_\do\%\do\~}
\def\mn@doi{\begingroup\mn@urlcharsother \@ifnextchar [ {\mn@doi@}
  {\mn@doi@[]}}
\def\mn@doi@[#1]#2{\def\@tempa{#1}\ifx\@tempa\@empty \href
  {http://dx.doi.org/#2} {doi:#2}\else \href {http://dx.doi.org/#2} {#1}\fi
  \endgroup}
\def\mn@eprint#1#2{\mn@eprint@#1:#2::\@nil}
\def\mn@eprint@arXiv#1{\href {http://arxiv.org/abs/#1} {{\tt arXiv:#1}}}
\def\mn@eprint@dblp#1{\href {http://dblp.uni-trier.de/rec/bibtex/#1.xml}
  {dblp:#1}}
\def\mn@eprint@#1:#2:#3:#4\@nil{\def\@tempa {#1}\def\@tempb {#2}\def\@tempc
  {#3}\ifx \@tempc \@empty \let \@tempc \@tempb \let \@tempb \@tempa \fi \ifx
  \@tempb \@empty \def\@tempb {arXiv}\fi \@ifundefined
  {mn@eprint@\@tempb}{\@tempb:\@tempc}{\expandafter \expandafter \csname
  mn@eprint@\@tempb\endcsname \expandafter{\@tempc}}}

\bibitem[\protect\citeauthoryear{{Auger}, {Treu}, {Bolton}, {Gavazzi},
  {Koopmans}, {Marshall}, {Moustakas}  \& {Burles}}{{Auger}
  et~al.}{2010}]{Auger_2010_fe}
{Auger} M.~W.,  {Treu} T.,  {Bolton} A.~S.,  {Gavazzi} R.,  {Koopmans}
  L.~V.~E.,  {Marshall} P.~J.,  {Moustakas} L.~A.,   {Burles} S.,  2010,
  \mn@doi [\apj] {10.1088/0004-637X/724/1/511}, \href
  {https://ui.adsabs.harvard.edu/abs/2010ApJ...724..511A} {724, 511}

\bibitem[\protect\citeauthoryear{{Barone-Nugent}, {Wyithe}, {Trenti}, {Treu},
  {Oesch}, {Bouwens}, {Illingworth}  \& {Schmidt}}{{Barone-Nugent}
  et~al.}{2015}]{Barone_Nugent_2015}
{Barone-Nugent} R.~L.,  {Wyithe} J.~S.~B.,  {Trenti} M.,  {Treu} T.,  {Oesch}
  P.,  {Bouwens} R.,  {Illingworth} G.~D.,   {Schmidt} K.~B.,  2015, \mn@doi
  [\mnras] {10.1093/mnras/stv633}, \href
  {https://ui.adsabs.harvard.edu/abs/2015MNRAS.450.1224B} {450, 1224}

\bibitem[\protect\citeauthoryear{{Bernardi}, {Shankar}, {Hyde}, {Mei},
  {Marulli}  \& {Sheth}}{{Bernardi} et~al.}{2010}]{Bernardi_VDF_2010}
{Bernardi} M.,  {Shankar} F.,  {Hyde} J.~B.,  {Mei} S.,  {Marulli} F.,
  {Sheth} R.~K.,  2010, \mn@doi [\mnras] {10.1111/j.1365-2966.2010.16425.x},
  \href {https://ui.adsabs.harvard.edu/abs/2010MNRAS.404.2087B} {404, 2087}

\bibitem[\protect\citeauthoryear{{Bezanson} et~al.,}{{Bezanson}
  et~al.}{2011}]{Bezanson_VDFevo}
{Bezanson} R.,  et~al., 2011, \mn@doi [\apjl] {10.1088/2041-8205/737/2/L31},
  \href {https://ui.adsabs.harvard.edu/abs/2011ApJ...737L..31B} {737, L31}

\bibitem[\protect\citeauthoryear{{Bolton}, {Burles}, {Koopmans}, {Treu},
  {Gavazzi}, {Moustakas}, {Wayth}  \& {Schlegel}}{{Bolton}
  et~al.}{2008a}]{Bolton_SLACS_2008}
{Bolton} A.~S.,  {Burles} S.,  {Koopmans} L. V.~E.,  {Treu} T.,  {Gavazzi} R.,
  {Moustakas} L.~A.,  {Wayth} R.,   {Schlegel} D.~J.,  2008a, \mn@doi [\apj]
  {10.1086/589327}, \href
  {https://ui.adsabs.harvard.edu/abs/2008ApJ...682..964B} {682, 964}

\bibitem[\protect\citeauthoryear{{Bolton}, {Treu}, {Koopmans}, {Gavazzi},
  {Moustakas}, {Burles}, {Schlegel}  \& {Wayth}}{{Bolton}
  et~al.}{2008b}]{Bolton_2008_fe}
{Bolton} A.~S.,  {Treu} T.,  {Koopmans} L. V.~E.,  {Gavazzi} R.,  {Moustakas}
  L.~A.,  {Burles} S.,  {Schlegel} D.~J.,   {Wayth} R.,  2008b, \mn@doi [\apj]
  {10.1086/589989}, \href
  {https://ui.adsabs.harvard.edu/abs/2008ApJ...684..248B} {684, 248}

\bibitem[\protect\citeauthoryear{{Bouwens}, {Illingworth}, {Ellis}, {Oesch}  \&
  {Stefanon}}{{Bouwens} et~al.}{2022}]{Bouwens21_data}
{Bouwens} R.~J.,  {Illingworth} G.,  {Ellis} R.~S.,  {Oesch} P.,   {Stefanon}
  M.,  2022, \mn@doi [\apj] {10.3847/1538-4357/ac86d1}, \href
  {https://ui.adsabs.harvard.edu/abs/2022ApJ...940...55B} {940, 55}

\bibitem[\protect\citeauthoryear{{Cabanac} et~al.,}{{Cabanac}
  et~al.}{2007}]{CFHTLS_SL2S}
{Cabanac} R.~A.,  et~al., 2007, \mn@doi [\aap] {10.1051/0004-6361:20065810},
  \href {https://ui.adsabs.harvard.edu/abs/2007A&A...461..813C} {461, 813}

\bibitem[\protect\citeauthoryear{{Capelo} \& {Natarajan}}{{Capelo} \&
  {Natarajan}}{2007}]{Capelo_Natarajan_2007}
{Capelo} P.~R.,  {Natarajan} P.,  2007, \mn@doi [New Journal of Physics]
  {10.1088/1367-2630/9/12/445}, \href
  {https://ui.adsabs.harvard.edu/abs/2007NJPh....9..445C} {9, 445}

\bibitem[\protect\citeauthoryear{{Chae}}{{Chae}}{2010}]{Chae_2010}
{Chae} K.-H.,  2010, \mn@doi [\mnras] {10.1111/j.1365-2966.2009.16073.x}, \href
  {https://ui.adsabs.harvard.edu/abs/2010MNRAS.402.2031C} {402, 2031}

\bibitem[\protect\citeauthoryear{{Chae} \& {Mao}}{{Chae} \&
  {Mao}}{2003}]{ChaeMao2003_VDFevo}
{Chae} K.-H.,  {Mao} S.,  2003, \mn@doi [\apjl] {10.1086/381247}, \href
  {https://ui.adsabs.harvard.edu/abs/2003ApJ...599L..61C} {599, L61}

\bibitem[\protect\citeauthoryear{{Choi}, {Park}  \& {Vogeley}}{{Choi}
  et~al.}{2007}]{Choi_2007_SDSS}
{Choi} Y.-Y.,  {Park} C.,   {Vogeley} M.~S.,  2007, \mn@doi [\apj]
  {10.1086/511060}, \href
  {https://ui.adsabs.harvard.edu/abs/2007ApJ...658..884C} {658, 884}

\bibitem[\protect\citeauthoryear{{Collett} et~al.,}{{Collett}
  et~al.}{2023}]{4SLSLS_COLLETT}
{Collett} T.~E.,  et~al., 2023, \mn@doi [The Messenger]
  {10.18727/0722-6691/5313}, \href
  {https://ui.adsabs.harvard.edu/abs/2023Msngr.190...49C} {190, 49}

\bibitem[\protect\citeauthoryear{{Correa}, {Wyithe}, {Schaye}  \&
  {Duffy}}{{Correa} et~al.}{2015}]{Correa_2015}
{Correa} C.~A.,  {Wyithe} J. S.~B.,  {Schaye} J.,   {Duffy} A.~R.,  2015,
  \mn@doi [\mnras] {10.1093/mnras/stv689}, \href
  {https://ui.adsabs.harvard.edu/abs/2015MNRAS.450.1514C} {450, 1514}

\bibitem[\protect\citeauthoryear{{Euclid Collaboration} et~al.,}{{Euclid
  Collaboration} et~al.}{2022}]{Euclid_Wide_Survey}
{Euclid Collaboration} et~al., 2022, \mn@doi [\aap]
  {10.1051/0004-6361/202141938}, \href
  {https://ui.adsabs.harvard.edu/abs/2022A&A...662A.112E} {662, A112}

\bibitem[\protect\citeauthoryear{{Faber} \& {Jackson}}{{Faber} \&
  {Jackson}}{1976}]{Faber_Jackson}
{Faber} S.~M.,  {Jackson} R.~E.,  1976, \mn@doi [\apj] {10.1086/154215}, \href
  {https://ui.adsabs.harvard.edu/abs/1976ApJ...204..668F} {204, 668}

\bibitem[\protect\citeauthoryear{{Ferrami} \& {Wyithe}}{{Ferrami} \&
  {Wyithe}}{2023}]{Ferrami_Lensing_bright_end}
{Ferrami} G.,  {Wyithe} J. S.~B.,  2023, \mn@doi [\mnras]
  {10.1093/mnrasl/slad050}, \href
  {https://ui.adsabs.harvard.edu/abs/2023MNRAS.523L..21F} {523, L21}

\bibitem[\protect\citeauthoryear{{Ferrami} \& {Wyithe}}{{Ferrami} \&
  {Wyithe}}{2024}]{Ferrami_Wyithe_Lens_stat_model}
{Ferrami} G.,  {Wyithe} J. S.~B.,  2024, \mn@doi [\mnras]
  {10.1093/mnras/stae1607}, \href
  {https://ui.adsabs.harvard.edu/abs/2024MNRAS.532.1832F} {532, 1832}

\bibitem[\protect\citeauthoryear{{Foreman-Mackey} et~al.,}{{Foreman-Mackey}
  et~al.}{2013}]{emcee}
{Foreman-Mackey} D.,  et~al., 2013, {emcee: The MCMC Hammer}, Astrophysics
  Source Code Library, record ascl:1303.002

\bibitem[\protect\citeauthoryear{{Gavazzi}, {Treu}, {Rhodes}, {Koopmans},
  {Bolton}, {Burles}, {Massey}  \& {Moustakas}}{{Gavazzi}
  et~al.}{2007}]{Gavazzi_SLACS_2007}
{Gavazzi} R.,  {Treu} T.,  {Rhodes} J.~D.,  {Koopmans} L. V.~E.,  {Bolton}
  A.~S.,  {Burles} S.,  {Massey} R.~J.,   {Moustakas} L.~A.,  2007, \mn@doi
  [\apj] {10.1086/519237}, \href
  {https://ui.adsabs.harvard.edu/abs/2007ApJ...667..176G} {667, 176}

\bibitem[\protect\citeauthoryear{{Geng}, {Cao}, {Liu}, {Liu}, {Biesiada}  \&
  {Lian}}{{Geng} et~al.}{2021}]{Geng_2021}
{Geng} S.,  {Cao} S.,  {Liu} Y.,  {Liu} T.,  {Biesiada} M.,   {Lian} Y.,  2021,
  \mn@doi [\mnras] {10.1093/mnras/stab519}, \href
  {https://ui.adsabs.harvard.edu/abs/2021MNRAS.503.1319G} {503, 1319}

\bibitem[\protect\citeauthoryear{{Grillo}, {Lombardi}  \& {Bertin}}{{Grillo}
  et~al.}{2008}]{Grillo_2008_fe}
{Grillo} C.,  {Lombardi} M.,   {Bertin} G.,  2008, \mn@doi [\aap]
  {10.1051/0004-6361:20077534}, \href
  {https://ui.adsabs.harvard.edu/abs/2008A&A...477..397G} {477, 397}

\bibitem[\protect\citeauthoryear{{Holloway}, {Verma}, {Marshall}, {More}  \&
  {Tecza}}{{Holloway} et~al.}{2023}]{Holloway_2023}
{Holloway} P.,  {Verma} A.,  {Marshall} P.~J.,  {More} A.,   {Tecza} M.,  2023,
  \mn@doi [\mnras] {10.1093/mnras/stad2371}, \href
  {https://ui.adsabs.harvard.edu/abs/2023MNRAS.525.2341H} {525, 2341}

\bibitem[\protect\citeauthoryear{{Ivezic} et~al.,}{{Ivezic}
  et~al.}{2008}]{Vera_Rubin_LSST}
{Ivezic} Z.,  et~al., 2008, \mn@doi [Serbian Astronomical Journal]
  {10.2298/SAJ0876001I}, \href
  {https://ui.adsabs.harvard.edu/abs/2008SerAJ.176....1I} {176, 1}

\bibitem[\protect\citeauthoryear{{Kochanek}}{{Kochanek}}{1992}]{Kochanek_1992}
{Kochanek} C.~S.,  1992, \mn@doi [\apj] {10.1086/170845}, \href
  {https://ui.adsabs.harvard.edu/abs/1992ApJ...384....1K} {384, 1}

\bibitem[\protect\citeauthoryear{{Kochanek}}{{Kochanek}}{1996}]{Kochanek_1996}
{Kochanek} C.~S.,  1996, \mn@doi [\apj] {10.1086/177538}, \href
  {https://ui.adsabs.harvard.edu/abs/1996ApJ...466..638K} {466, 638}

\bibitem[\protect\citeauthoryear{{Kochanek} et~al.,}{{Kochanek}
  et~al.}{2000}]{Kochanek_2000_fe}
{Kochanek} C.~S.,  et~al., 2000, \mn@doi [\apj] {10.1086/317074}, \href
  {https://ui.adsabs.harvard.edu/abs/2000ApJ...543..131K} {543, 131}

\bibitem[\protect\citeauthoryear{{Koopmans} et~al.,}{{Koopmans}
  et~al.}{2009}]{Koopmans_bulge_halo_conspiracy}
{Koopmans} L.~V.~E.,  et~al., 2009, \mn@doi [\apjl]
  {10.1088/0004-637X/703/1/L51}, \href
  {https://ui.adsabs.harvard.edu/abs/2009ApJ...703L..51K} {703, L51}

\bibitem[\protect\citeauthoryear{{Lapi}, {Negrello}, {Gonz{\'a}lez-Nuevo},
  {Cai}, {De Zotti}  \& {Danese}}{{Lapi} et~al.}{2012}]{Lapi_2012}
{Lapi} A.,  {Negrello} M.,  {Gonz{\'a}lez-Nuevo} J.,  {Cai} Z.~Y.,  {De Zotti}
  G.,   {Danese} L.,  2012, \mn@doi [\apj] {10.1088/0004-637X/755/1/46}, \href
  {https://ui.adsabs.harvard.edu/abs/2012ApJ...755...46L} {755, 46}

\bibitem[\protect\citeauthoryear{{Mason} et~al.,}{{Mason}
  et~al.}{2015}]{Mason_2015}
{Mason} C.~A.,  et~al., 2015, \mn@doi [\apj] {10.1088/0004-637X/805/1/79},
  \href {https://ui.adsabs.harvard.edu/abs/2015ApJ...805...79M} {805, 79}

\bibitem[\protect\citeauthoryear{{Matsumoto} \& {Futamase}}{{Matsumoto} \&
  {Futamase}}{2008}]{Matsumoto_Futamase_2008}
{Matsumoto} A.,  {Futamase} T.,  2008, \mn@doi [\mnras]
  {10.1111/j.1365-2966.2007.12769.x}, \href
  {https://ui.adsabs.harvard.edu/abs/2008MNRAS.384..843M} {384, 843}

\bibitem[\protect\citeauthoryear{{Mitchell}, {Keeton}, {Frieman}  \&
  {Sheth}}{{Mitchell} et~al.}{2005}]{Mitchell_2005_VDF_from_LF}
{Mitchell} J.~L.,  {Keeton} C.~R.,  {Frieman} J.~A.,   {Sheth} R.~K.,  2005,
  \mn@doi [\apj] {10.1086/427910}, \href
  {https://ui.adsabs.harvard.edu/abs/2005ApJ...622...81M} {622, 81}

\bibitem[\protect\citeauthoryear{{Montero-Dorta}, {Bolton}  \&
  {Shu}}{{Montero-Dorta} et~al.}{2017}]{Montero_Dorta_VDF}
{Montero-Dorta} A.~D.,  {Bolton} A.~S.,   {Shu} Y.,  2017, \mn@doi [\mnras]
  {10.1093/mnras/stx321}, \href
  {https://ui.adsabs.harvard.edu/abs/2017MNRAS.468...47M} {468, 47}

\bibitem[\protect\citeauthoryear{{Ofek}, {Rix}  \& {Maoz}}{{Ofek}
  et~al.}{2003}]{Ofek_2003}
{Ofek} E.~O.,  {Rix} H.-W.,   {Maoz} D.,  2003, \mn@doi [\mnras]
  {10.1046/j.1365-8711.2003.06707.x}, \href
  {https://ui.adsabs.harvard.edu/abs/2003MNRAS.343..639O} {343, 639}

\bibitem[\protect\citeauthoryear{{Oguri} \& {Marshall}}{{Oguri} \&
  {Marshall}}{2010}]{Oguri_Marshall}
{Oguri} M.,  {Marshall} P.~J.,  2010, \mn@doi [\mnras]
  {10.1111/j.1365-2966.2010.16639.x}, \href
  {https://ui.adsabs.harvard.edu/abs/2010MNRAS.405.2579O} {405, 2579}

\bibitem[\protect\citeauthoryear{{Oguri}, {Keeton}  \& {Dalal}}{{Oguri}
  et~al.}{2005}]{Oguri2005}
{Oguri} M.,  {Keeton} C.~R.,   {Dalal} N.,  2005, \mn@doi [\mnras]
  {10.1111/j.1365-2966.2005.09697.x}, \href
  {https://ui.adsabs.harvard.edu/abs/2005MNRAS.364.1451O} {364, 1451}

\bibitem[\protect\citeauthoryear{{Oguri} et~al.,}{{Oguri}
  et~al.}{2012a}]{OguriVDF2012}
{Oguri} M.,  et~al., 2012a, \mn@doi [\aj] {10.1088/0004-6256/143/5/120}, \href
  {https://ui.adsabs.harvard.edu/abs/2012AJ....143..120O} {143, 120}

\bibitem[\protect\citeauthoryear{{Oguri} et~al.,}{{Oguri}
  et~al.}{2012b}]{Oguri_2012}
{Oguri} M.,  et~al., 2012b, \mn@doi [\aj] {10.1088/0004-6256/143/5/120}, \href
  {https://ui.adsabs.harvard.edu/abs/2012AJ....143..120O} {143, 120}

\bibitem[\protect\citeauthoryear{{Shankar}, {Bernardi}  \& {Haiman}}{{Shankar}
  et~al.}{2009}]{2009Shankar}
{Shankar} F.,  {Bernardi} M.,   {Haiman} Z.,  2009, \mn@doi [\apj]
  {10.1088/0004-637X/694/2/867}, \href
  {https://ui.adsabs.harvard.edu/abs/2009ApJ...694..867S} {694, 867}

\bibitem[\protect\citeauthoryear{{Sheth} et~al.,}{{Sheth}
  et~al.}{2003}]{Sheth_2003_VDF_from_LF}
{Sheth} R.~K.,  et~al., 2003, \mn@doi [\apj] {10.1086/376794}, \href
  {https://ui.adsabs.harvard.edu/abs/2003ApJ...594..225S} {594, 225}

\bibitem[\protect\citeauthoryear{{Shibuya}, {Ouchi}  \& {Harikane}}{{Shibuya}
  et~al.}{2015}]{Shibuya_L_size_rel}
{Shibuya} T.,  {Ouchi} M.,   {Harikane} Y.,  2015, \mn@doi [\apjs]
  {10.1088/0067-0049/219/2/15}, \href
  {https://ui.adsabs.harvard.edu/abs/2015ApJS..219...15S} {219, 15}

\bibitem[\protect\citeauthoryear{{Sohn}, {Zahid}  \& {Geller}}{{Sohn}
  et~al.}{2017}]{Sohn_VDF_2017}
{Sohn} J.,  {Zahid} H.~J.,   {Geller} M.~J.,  2017, \mn@doi [\apj]
  {10.3847/1538-4357/aa7de3}, \href
  {https://ui.adsabs.harvard.edu/abs/2017ApJ...845...73S} {845, 73}

\bibitem[\protect\citeauthoryear{{Sonnenfeld}}{{Sonnenfeld}}{2024}]{SLACS_debiased}
{Sonnenfeld} A.,  2024, \mn@doi [\aap] {10.1051/0004-6361/202451341}, \href
  {https://ui.adsabs.harvard.edu/abs/2024A&A...690A.325S} {690, A325}

\bibitem[\protect\citeauthoryear{{Sonnenfeld}, {Gavazzi}, {Suyu}, {Treu}  \&
  {Marshall}}{{Sonnenfeld} et~al.}{2013a}]{Sonnenfeld_SL2S_2013a}
{Sonnenfeld} A.,  {Gavazzi} R.,  {Suyu} S.~H.,  {Treu} T.,   {Marshall} P.~J.,
  2013a, \mn@doi [\apj] {10.1088/0004-637X/777/2/97}, \href
  {https://ui.adsabs.harvard.edu/abs/2013ApJ...777...97S} {777, 97}

\bibitem[\protect\citeauthoryear{{Sonnenfeld}, {Treu}, {Gavazzi}, {Suyu},
  {Marshall}, {Auger}  \& {Nipoti}}{{Sonnenfeld}
  et~al.}{2013b}]{Sonnenfeld_SL2S_2013}
{Sonnenfeld} A.,  {Treu} T.,  {Gavazzi} R.,  {Suyu} S.~H.,  {Marshall} P.~J.,
  {Auger} M.~W.,   {Nipoti} C.,  2013b, \mn@doi [\apj]
  {10.1088/0004-637X/777/2/98}, \href
  {https://ui.adsabs.harvard.edu/abs/2013ApJ...777...98S} {777, 98}

\bibitem[\protect\citeauthoryear{{Spergel} et~al.,}{{Spergel}
  et~al.}{2015}]{Roman_WFIRST}
{Spergel} D.,  et~al., 2015, \mn@doi [arXiv e-prints]
  {10.48550/arXiv.1503.03757}, \href
  {https://ui.adsabs.harvard.edu/abs/2015arXiv150303757S} {p. arXiv:1503.03757}

\bibitem[\protect\citeauthoryear{Stockmann et~al.,}{Stockmann
  et~al.}{2021}]{Stockmann_2021}
Stockmann M.,  et~al., 2021, \mn@doi [The Astrophysical Journal]
  {10.3847/1538-4357/abce66}, 908, 135

\bibitem[\protect\citeauthoryear{{Tan} et~al.,}{{Tan}
  et~al.}{2024}]{ProjectDinos1}
{Tan} C.~Y.,  et~al., 2024, \mn@doi [\mnras] {10.1093/mnras/stae884}, \href
  {https://ui.adsabs.harvard.edu/abs/2024MNRAS.530.1474T} {530, 1474}

\bibitem[\protect\citeauthoryear{{Taylor} et~al.,}{{Taylor}
  et~al.}{2022}]{Taylor_VDF}
{Taylor} L.,  et~al., 2022, \mn@doi [\apj] {10.3847/1538-4357/ac9796}, \href
  {https://ui.adsabs.harvard.edu/abs/2022ApJ...939...90T} {939, 90}

\bibitem[\protect\citeauthoryear{{Treu}, {Koopmans}, {Bolton}, {Burles}  \&
  {Moustakas}}{{Treu} et~al.}{2006}]{Treu_2006}
{Treu} T.,  {Koopmans} L.~V.,  {Bolton} A.~S.,  {Burles} S.,   {Moustakas}
  L.~A.,  2006, \mn@doi [\apj] {10.1086/500124}, \href
  {https://ui.adsabs.harvard.edu/abs/2006ApJ...640..662T} {640, 662}

\bibitem[\protect\citeauthoryear{{Treu}, {Gavazzi}, {Gorecki}, {Marshall},
  {Koopmans}, {Bolton}, {Moustakas}  \& {Burles}}{{Treu}
  et~al.}{2009}]{SLACSVIII_Treu}
{Treu} T.,  {Gavazzi} R.,  {Gorecki} A.,  {Marshall} P.~J.,  {Koopmans} L.
  V.~E.,  {Bolton} A.~S.,  {Moustakas} L.~A.,   {Burles} S.,  2009, \mn@doi
  [\apj] {10.1088/0004-637X/690/1/670}, \href
  {https://ui.adsabs.harvard.edu/abs/2009ApJ...690..670T} {690, 670}

\bibitem[\protect\citeauthoryear{{Tully} \& {Fisher}}{{Tully} \&
  {Fisher}}{1977}]{Tully_Fisher}
{Tully} R.~B.,  {Fisher} J.~R.,  1977, \aap, \href
  {https://ui.adsabs.harvard.edu/abs/1977A&A....54..661T} {54, 661}

\bibitem[\protect\citeauthoryear{{Turner}, {Ostriker}  \& {Gott}}{{Turner}
  et~al.}{1984}]{Turner_1984}
{Turner} E.~L.,  {Ostriker} J.~P.,   {Gott} J.~R. I.,  1984, \mn@doi [\apj]
  {10.1086/162379}, \href
  {https://ui.adsabs.harvard.edu/abs/1984ApJ...284....1T} {284, 1}

\bibitem[\protect\citeauthoryear{{Zahid}, {Sohn}  \& {Geller}}{{Zahid}
  et~al.}{2018}]{Zahid_2018_fe}
{Zahid} H.~J.,  {Sohn} J.,   {Geller} M.~J.,  2018, \mn@doi [\apj]
  {10.3847/1538-4357/aabe31}, \href
  {https://ui.adsabs.harvard.edu/abs/2018ApJ...859...96Z} {859, 96}

\bibitem[\protect\citeauthoryear{{van der Wel} et~al.,}{{van der Wel}
  et~al.}{2014}]{van_der_Wel_2014}
{van der Wel} A.,  et~al., 2014, \mn@doi [\apjl] {10.1088/2041-8205/792/1/L6},
  \href {https://ui.adsabs.harvard.edu/abs/2014ApJ...792L...6V} {792, L6}

\makeatother
\end{thebibliography}

% \appendix
% In this Appendix we show how we obtained the interval of 

% \section{Accounting for limits to ground based NIR spec-z}\label{sec:Appendix}
% \begin{figure*}
%   \includegraphics[width=\linewidth]{img/NIR_spec_limits.png}
%     \caption{NIR spec limits}
%     \label{fig:NIR_spec_limits}
% \end{figure*}

% Don't change these lines
\bsp  % typesetting comment
\label{lastpage}
\end{document}